\documentclass[aps,prd,preprintnumbers,groupedaddress,nofootinbib,amssymb,eqsecnum,notitlepage]{revtex4}
\usepackage{graphicx}
\usepackage{bm}
\usepackage{amsmath}
\usepackage{color}
\usepackage{amsfonts}
\usepackage{here}
\usepackage{graphicx}
\usepackage{amsmath,amsthm,amssymb}
\usepackage{bm}
\allowdisplaybreaks[1]

\usepackage{amsfonts}
\usepackage{dcolumn}
\usepackage{hyperref}

\begin{document}
\newcommand{\newc}{\newcommand}

\newcommand{\ben}{\begin{eqnarray}}
\newcommand{\een}{\end{eqnarray}}
\newc{\be}{\begin{equation}}
\newc{\ee}{\end{equation}}
\newc{\ba}{\begin{eqnarray}}
\newc{\ea}{\end{eqnarray}}
\newc{\bea}{\begin{eqnarray*}}
\newc{\eea}{\end{eqnarray*}}
\newc{\D}{\partial}
\newc{\ie}{{\it i.e.} }
\newc{\eg}{{\it e.g.} }
\newc{\etc}{{\it etc.} }
\newc{\etal}{{\it et al.}}
\newcommand{\nn}{\nonumber}
\newc{\ra}{\rightarrow}
\newc{\lra}{\leftrightarrow}
\newc{\lsim}{\buildrel{<}\over{\sim}}
\newc{\gsim}{\buildrel{>}\over{\sim}}
\newc{\aP}{\alpha_{\rm P}}

\title{Black holes in vector-tensor theories}

\author{
Lavinia Heisenberg$^{1}$, 
Ryotaro Kase$^{2}$,
Masato Minamitsuji$^{3}$, and 
Shinji Tsujikawa$^{2}$}

\affiliation{
$^1$Institute for Theoretical Studies, ETH Zurich, Clausiusstrasse 47, 8092 Zurich, Switzerland\\
$^2$Department of Physics, Faculty of Science, Tokyo University of Science, 1-3, Kagurazaka,
Shinjuku-ku, Tokyo 162-8601, Japan\\
$^3$Centro Multidisciplinar de Astrofisica - CENTRA, Departamento 
de Fisica, Instituto Superior Tecnico - IST, Universidade de Lisboa - 
UL, Avenida Rovisco Pais 1, 1049-001 Lisboa, Portugal}

\date{\today}

\begin{abstract}
We study static and spherically symmetric black hole (BH) solutions in second-order generalized Proca theories with nonminimal vector field derivative couplings to the Ricci scalar, the Einstein tensor, and the double dual Riemann tensor.
We find concrete Lagrangians which give rise to exact BH solutions 
by imposing two conditions of the two identical metric components 
and the constant norm of the vector field. 
These exact solutions are described by either Reissner-Nordstr\"{o}m (RN), stealth Schwarzschild, or extremal RN solutions
with a non-trivial longitudinal mode of the vector field.
We then numerically construct BH solutions without imposing these conditions. For cubic and quartic Lagrangians with power-law couplings 
which encompass vector Galileons as the specific cases, 
we show the existence of BH solutions 
with the difference between two non-trivial metric components.
The quintic-order power-law couplings do not give rise to non-trivial 
BH solutions regular throughout the horizon exterior.
The sixth-order and intrinsic vector-mode couplings can lead to 
BH solutions with a secondary hair. 
For all the solutions, the vector field is regular at least at the future 
or past horizon.
The deviation from General Relativity induced 
by the Proca hair can be potentially tested by future measurements of 
gravitational waves in the nonlinear regime of gravity.
\end{abstract}


\maketitle

\section{Introduction}
\label{introsec}

The direct detection of gravitational waves by 
Advanced LIGO \cite{LIGO} opened up a new opportunity for 
probing the physics of black holes (BHs) and their surroundings. 
The Solar System experiments of gravity 
have shown that General Relativity (GR) holds in high precision 
in the weak gravitational regime of the local Universe \cite{Will}.
The future gravitational wave experiments will allow us to 
test the validity of GR and possible deviations from GR 
in the nonlinear regime of gravity \cite{Vitor,Yagi}. 

{}From the cosmological side, the observational evidence of 
dark energy and dark matter \cite{SNIa,CMB} implies 
that there may be some extra propagating 
degrees of freedom (DOFs) beyond the realm of GR. 
In particular, the infrared modifications of gravity with new DOFs 
have been extensively studied as sources for the late-time cosmic 
acceleration \cite{moreview}. 
In the local Universe with weak gravitational 
backgrounds, the propagation of fifth forces mediated 
by new DOFs can be suppressed 
under the operation of screening mechanisms, e.g., 
Vainshtein \cite{Vain} or chameleon \cite{chame} mechanisms.
In the strong-gravity regime associated with 
BHs and neutron stars (NSs), the behavior of new DOFs is generally more involved due to tensor nonlinearities.

In the Einstein-Maxwell system of GR without matter, there is a uniqueness theorem 
stating that the asymptotically flat and stationary BH solutions are described 
only by three parameters, i.e., mass, electric charge, and angular 
momentum \cite{Israel,Carter,Wheeler,Hawking}. 
The ``no-hair'' BH theorem is valid for a canonical scalar field 
minimally coupled to gravity \cite{Chase,BekenPRL}. 
The same no-hair property also persists for
standard scalar-tensor theories in which the field has a direct 
coupling to the Ricci scalar \cite{Hawking72,Beken95,Soti12}.
However, the no-hair theorem of Ref.~\cite{Beken95} loses its validity 
in modified gravitational theories with nonminimal 
derivative couplings to gravity. 
The typical examples of such derivative couplings are 
Galileons \cite{Galileon1,Galileon2}, 
whose equations of motion respect the Galilean symmetry in the 
Minkowski limit. The extension of Galileons to more general couplings 
led to the rediscovery of Horndeski 
theories \cite{redis}--most general scalar-tensor 
theories with second-order equations of motion \cite{Horndeski}.

In the case of shift-symmetric Horndeski theories including Galileons, 
Hui and Nicolis \cite{Hui} argued conditions for the no-hair properties of BHs 
by utilizing the properties of the conserved
Noether current $J^\mu$. The existence of the shift symmetry gives rise to the 
field equation $\nabla_{\mu}J^{\mu}=0$, where $\nabla_{\mu}$ denotes 
the covariant derivative.
Under the assumptions that 
i) the spacetime is static, spherically-symmetric and 
asymptotically flat,
ii) the scalar field $\phi$ respects the symmetry of spacetime,
i.e., $\phi=\phi(r)$, where $r$ is the distance from the center of 
symmetry, and 
iii) the scalar product $J_\mu J^\mu$ is regular everywhere,
it can be shown that the radial current $J^r$ needs to 
vanish at all the distance $r$ due to the regularity 
on the horizon,  $J^r=0$.
They further employed the fact that 
the current takes the form 
$J^r=\phi' g^{rr}{\cal F}(\phi';g,g',g'')$, where 
$\phi'=d\phi/dr$, 
$g^{rr}$ is the radial component of the metric $g^{\mu \nu}$, 
and ${\cal F}$ is a function containing $\phi'$ and derivatives of $g^{\mu \nu}$.
Provided ${\cal F}$ does not contain negative powers 
of $\phi'$ so that the canonical kinetic term dominates in the asymptotically flat region, $g^{rr}$ and ${\cal F}$ approach non-vanishing constant values
and hence $\phi'=0$ at the infinity.
Moving inward from infinity, 
$g^{rr}$ and ${\cal F}$ vary continuously taking non-zero values,
so only the allowed profile consistent with $J^r=0$
is the no-hair solution satisfying $\phi'=0$.
In other words, in order to have a non-trivial hairy BH solution,
one has to break at least one of the assumptions made 
by Hui and Nicolis.

If the constancy of ${\cal F}$ is not imposed in the limit $\phi' \to 0$, 
there exist some hairy BH solutions in shift-symmetric 
Horndeski theories with non-vanishing values of $\phi'$
\cite{Rinaldi,Anabalon,Minami13,Soti1,Soti2,Babi17}. 
In the case where the scalar field is linearly 
coupled to a Gauss-Bonnet term,
the function ${\cal F}$ contains a negative power of $\phi'$,
which allows 
the existence of solutions with $\phi' \neq 0$ \cite{Soti1,Soti2}.  
Another approach to construct BH solutions in Horndeski theories
is to assume a linearly time-dependent scalar field
$\phi=qt+\psi (r)$ \cite{Babi14},
where ${\cal F}=0$ 
is ensured as the consequence of the equations of motion,
leaving $\phi'$ unfixed by the condition $J^r=0$.
This leads to a family of BH solutions with 
a static metric \cite{Koba14,Babi16,Lefteris}, 
especially the stealth Schwarzschild BH solution \cite{Babi14}.

If we consider a massless vector field $A_{\mu}$ with 
the Lagrangian $F=-F_{\mu \nu}F^{\mu \nu}/4$, where 
$F_{\mu \nu}=\nabla_{\mu}A_{\nu}-\nabla_{\nu}A_{\mu}$ is 
the field strength tensor, the resulting static and spherically symmetric 
solution in GR is given by the Reissner-Nordstr\"{o}m (RN) metric with mass $M$ and electric charge $Q$. 
For a massive vector field with the Lagrangian 
$-m^2 A^\mu A_\mu/2$, the $U(1)$ gauge symmetry is explicitly 
broken, so there is a longitudinal propagation besides two transverse 
polarizations. In the framework of GR, Bekenstein \cite{Beken72} showed that this massive Proca field $A_{\mu}$ needs to vanish due to the regularity of a physical scalar constructed from $A_{\mu}$ on the horizon. Hence the resulting solution is given by the 
Schwarzschild solution without the vector hair. 

This no-hair theorem for the massive Proca field cannot be applied to theories with vector derivative couplings. 
The action of generalized Proca theories with nonminimal derivative couplings to gravity 
was first constructed in Refs.~\cite{Heisenberg,Tasinato} from the demand of 
keeping the three propagating DOFs besides two tensor polarizations. The theories can be further extended \cite{Allys} to 
include intrinsic vector-mode couplings with the double 
dual Riemann tensor $L^{\mu \nu \alpha \beta}$ \cite{Jimenez2016}, such that the $U(1)$-invariant interactions derived by Horndeski \cite{Horndeski76} can be accommodated as a specific case. The equations of motion 
in these theories remain of second order, but one can build more general vector-tensor interactions beyond the second-order domain without introducing extra DOFs associated with the 
Ostrogradski instability \cite{HKT,Kimura}. 
In second-order generalized Proca theories and their 
extensions the derivative interactions can 
drive the late-time cosmic acceleration \cite{DeFelicecosmo} with some distinct observational 
signatures \cite{obsig1,obsig2}, while satisfying 
local gravity constraints in the Solar System \cite{DeFeliceVain}. 
See Ref.~\cite{Heisenberg:2017mzp} for a short summary. 

The study of hairy BH solutions in generalized Proca theories  
with an Abelian vector field $A_{\mu}$ has recently 
received attention for probing physics in the nonlinear regime 
of gravity \cite{Chagoya,Fan,Minami,Cisterna,Chagoya2,Babichev17,HKMT},
see also Refs.~\cite{colored} for early works of 
BH solutions in the presence of non-Abelian Yangs-Mills fields. 
In theories whose Lagrangians contain the coupling 
$\beta_4 G^{\mu \nu}
A_{\mu}A_{\nu}$ as well as $-F^{\mu \nu}F_{\mu \nu}/4$ 
and the Einstein-Hilbert term, Chagoya {\it et al.} \cite{Chagoya} derived an exact spherically symmetric and static BH solution for the 
specific coupling $\beta_4=1/4$.
This exact BH solution was further extended 
to asymptotically non-flat solutions \cite{Minami,Babichev17}, 
non-exact solutions for $\beta_4 \neq 1/4$ \cite{Babichev17,Chagoya2}, rotating solutions \cite{Minami}, 
and NSs \cite{Chagoya2}.

On a static and spherically symmetric background with the 
radial coordinate $r$, the vector field is characterized by the temporal 
component $A_0(r)$ and the longitudinal mode $A_1(r)$.
In general, the equation of motion for $A_1$
can be written in the form ${\cal F}(A_1,A_0,A_0';g,g')=0$. 
Unlike scalar-tensor theories, the presence of two vector 
components naturally allows the solution $A_1 \neq 0$
without restricting the functional form of ${\cal F}$. 
Hence it is not difficult to find hairy BH solutions even 
for simple power-law couplings like those of vector Galileons \cite{HKMT}.
For some derivative interactions the equation for $A_1$ 
reduces to the form $A_1 \tilde{\cal F}(A_1,A_0,A_0';g,g')=0$, 
so there is the branch $A_1=0$ besides $\tilde{\cal F}=0$.
Even with the branch $A_1=0$, the deviation from GR can arise in the metric components due to a modification of the temporal component 
$A_0$ induced by derivative couplings, e.g., 
the BH solution arising from the $U(1)$-invariant 
interaction \cite{HorndeskiBH}.

In this paper, we will present a detailed study of BH solutions 
in second-order generalized Proca theories 
by extending the analysis of Ref. \cite{HKMT}.
We will consider the full set of Lagrangians 
${\cal L}_{2,3,4,5,6}$ of the generalized Proca theories \cite{Heisenberg, Jimenez2016}.
Our analysis also covers the generalized quadratic-order Lagrangian $G_2(X,F,Y)$,
where $X=-A_{\mu}A^{\mu}/2$ and $Y=A^{\mu}A^{\nu}{F_{\mu}}^{\alpha}F_{\nu \alpha}$.
We also explain in details how to construct non-exact BH solutions in power-law coupling models 
containing the dependence of $X^n$ in each Lagrangian, 
where $n$ is a positive integer.

We organize our paper as follows.
In Sec.~\ref{modelsec} we present 
the full equations of motion on a static and spherically 
symmetric background and revisit the 
Bekenstein's no-hair BH solution for a massive Proca field.
In Sec.~\ref{exactsec} we review the exact BH solution
present for the quartic derivative coupling $G_4(X)$.
In Sec.~\ref{exactsec2} we construct a family of exact 
BH solutions in the presence of other couplings under 
the conditions that the two
metric components
are identical and that the norm $X$ of the vector field is constant. 
In Sec.~\ref{G3powersec} we study the BH solutions for 
cubic-order power-law coupling models $G_3(X) \propto X^n$ 
including vector Galileons ($n=1$) and numerically confirm 
the existence of regular hairy BH solutions outside the horizon.
Similarly,
in Secs.~\ref{G4powersec}, \ref{G5powersec}, \ref{G6powersec},
\ref{g45powersec} we
clarify the cases in which the BH 
solutions with primary or secondary Proca hairs are present for 
the power-law models containing the 
$X^n$ dependence in the couplings 
$G_4(X), G_5(X)$ and the intrinsic vector-mode couplings 
$G_6(X), g_{4,5}(X)$, respectively.
The last Sec.~\ref{concludesec} is devoted to conclusions.

\section{Generalized Proca theories}
\label{modelsec}

In the presence of a vector field $A_{\mu}$ with the field strength tensor 
$F_{\mu \nu}=\nabla_{\mu}A_{\nu}-\nabla_{\nu}A_{\mu}$, we consider second-order 
generalized Proca theories given by the action 
\be
S=\int d^{4}x \sqrt{-g} 
\left( F+\sum_{i=2}^{6} \mathcal{L}_{i} \right)\,,
\label{action}
\ee
where $F=-F_{\mu \nu}F^{\mu \nu}/4$ is the standard 
Maxwell term, $g$ is a 
determinant of the metric tensor $g_{\mu\nu}$, and \cite{Heisenberg, Jimenez2016}
\ba
\mathcal{L}_{2}&=& G_{2}(X, F, Y)\,,
\label{L2}
\\
\mathcal{L}_{3}&=& G_{3}(X) \nabla_{\mu} A^{\mu}\,,
\label{L3}
\\
\mathcal{L}_{4}&=& G_{4}(X) R 
+ G_{4,X}(X)\left[ (\nabla_{\mu} A^{\mu})^{2} 
-  \nabla_{\mu} A_{\nu} \nabla^{\nu} A^{\mu}\right] \,,\\
\mathcal{L}_{5}&=& G_{5}(X) G_{\mu\nu} \nabla^{\mu} A^{\nu} 
- \frac{1}{6} G_{5,X} (X) \left[ (\nabla_{\mu} A^{\mu})^{3} 
- 3 \nabla_{\mu} A^{\mu} \nabla_{\rho} A_{\sigma} \nabla^{\sigma} A^{\rho} 
+ 2 \nabla_{\rho} A_{\sigma} \nabla^{\nu} A^{\rho} \nabla^{\sigma} A_{\nu} \right]
\nn
\\
&&-g_{5}(X) \tilde{F}^{\alpha\mu}\tilde{F}^{\beta}_{~\mu} \nabla_{\alpha} A_{\beta}\,,
\\
\mathcal{L}_{6}&=& G_{6}(X) L^{\mu\nu\alpha\beta} \nabla_{\mu} A_{\nu} \nabla_{\alpha} A_{\beta}
+\frac{1}{2} G_{6,X}(X) \tilde{F}^{\alpha\beta} \tilde{F}^{\mu\nu} \nabla_{\alpha} 
A_{\mu} \nabla_{\beta} A_{\nu}\,,
\label{L6}
\ea
with $X=-A_{\mu}A^{\mu}/2$, $Y=A^{\mu}A^{\nu}{F_{\mu}}^{\alpha}F_{\nu \alpha}$ and 
$G_{i,X}=\partial G_i/\partial X$.
The functions $G_{3,4,5,6}$ and $g_5$ depend on 
$X$ alone, whereas the function 
$G_{2}$ is dependent on $F$ as well as $X$ and $Y$. 
The quantity $\tilde{F}^{\mu\nu}$ is the dual strength 
tensor given by $\tilde{F}^{\mu\nu}=\mathcal{E}^{\mu\nu\alpha\beta} F_{\alpha\beta}/2$, 
where $\mathcal{E}^{\mu\nu\alpha\beta}$ is the Levi-Civita tensor 
satisfying the normalization  
$\mathcal{E}^{\mu\nu\alpha\beta}\mathcal{E}_{\mu\nu\alpha\beta}=-4!$. The vector field has derivative couplings to 
the Ricci scalar $R$, the Einstein tensor $G_{\mu \nu}$, 
and the double dual Riemann tensor 
$L^{\mu \nu \alpha \beta}$ defined by 
\be
L^{\mu\nu\alpha\beta}=\frac{1}{4} \mathcal{E}^{\mu\nu\rho\sigma} \mathcal{E}^{\alpha\beta\gamma\delta} R_{\rho\sigma\gamma\delta}\,,
\ee
where $R_{\rho\sigma\gamma\delta}$ is the Riemann tensor.

The original Proca theory with the mass term $m$ corresponds 
to the Lagrangian $G_2=m^2X$, in which case the longitudinal propagation arises besides two transverse 
polarizations. The action (\ref{action}) has been constructed 
to keep the propagating DOFs of scalar and vector modes 
unchanged, i.e., three DOFs. 
Taking the scalar limit $A_{\mu} \to \nabla_{\mu} \pi$, 
the quantity $F$ as well as the Lagrangians $-g_{5}(X) \tilde{F}^{\alpha\mu}\tilde{F}^{\beta}_{~\mu} \nabla_{\alpha} A_{\beta}$ and ${\cal L}_6$ vanish, so they correspond to 
intrinsic vector modes. In Ref.~\cite{HKMT} there is 
the term $-2g_4(X)F$ in ${\cal L}_4$, but such a term is 
now absorbed into the Lagrangian ${\cal L}_2=G_2(X, F, Y)$. 

To study BH solutions on a static and spherically symmetric 
background, we take the line element 
\be
ds^{2} =-f(r) dt^{2} +h^{-1}(r)dr^{2} + 
r^{2} (d\theta^{2}+\sin^{2}\theta\, d \varphi^{2}),
\label{metric}
\ee
where 
$t$, $r$ and $(\theta,\varphi)$ represent the time, radial, and angular coordinates, respectively,
$f(r)$ and $h(r)$ are functions of $r$
such that $f(r)>0$ and $h(r)>0$ outside the event horizon $r>r_h$,
$r_h$ is the position of the horizon at which $f(r_h)=h(r_h)=0$. 
Expressing the vector field in the form 
$A_{\mu}=(A_0,A_i)$, the spatial vector $A_i$ 
can be decomposed into the transverse and longitudinal 
components, as $A_i=A_i^{(T)}+\nabla_i \chi$, where 
$A_i^{(T)}$ obeys the transverse condition 
$\nabla^{i}A_i^{(T)}=0$ and $\chi$ is the longitudinal 
scalar. {}From the regularity at the origin of the static and spherically symmetric 
background the transverse mode 
$A_i^{(T)}$ needs to vanish \cite{DeFeliceVain}, 
so we are left with the longitudinal scalar $\chi$ in $A_i$. 
Hence the vector-field profile compatible with 
the background (\ref{metric}) is given by 
\be
A_{\mu}=\left( A_0(r), A_1(r), 0, 0 \right)\,,
\label{vector_ansatz}
\ee
where $A_1(r)=\chi'(r)$, and a prime represents the 
derivative with respect to $r$.

At this stage, we would like to make a comment about the additional intrinsic vector-mode contribution 
$Y$ in  ${\cal L}_2= G_2(X,F,Y)$.
On the static and spherically symmetric background (\ref{metric}) 
with the vector components (\ref{vector_ansatz}) we have that 
$Y=4FX$, so the additional dependence of $Y$ in Eq.~(\ref{L2}) 
can be removed. Thus, we will work on the quadratic-order Lagrangian 
\be
{\cal L}_2= G_2(X,F)\,,
\ee
in the rest of the paper.

The term $X=-A_{\mu}A^{\mu}/2$ can be expressed as 
$X=X_{0}+X_{1}$, where 
\be
X_{0}=\frac{A_{0}^{2}}{2 f}, \qquad
X_{1}=-\frac{h A_{1}^{2}}{2}.
\ee
Varying the action (\ref{action}) with respect to $A_0$ and 
$A_1$ respectively gives rise to the vector-field equations. 
The equation of motion for $A_0$ results in
\ba
\hspace{-0.4cm}
& & 
rf \left[ 2fh(rA_0''+2A_0')+r(fh'-f'h)A_0' \right] (1+G_{2,F})
+r^2hA_0'^2 \left[ 2fhA_0''-(f'h-fh') A_0' \right] G_{2,FF}
-2 r^2f^2A_0 G_{2,X}
\notag\\
\hspace{-0.4cm}
& & 
-2 r^2fA_0' \left( fh^2A_1A_1' -hA_0A_0'+f'hX_0-fh'X_1\right) G_{2,XF}
-rfA_0 \left[ 2 rfhA_1'+(rf'h+rfh'+4fh)A_1 \right] G_{3,X}
\notag\\
\hspace{-0.4cm}
& & 
+4 f^2A_0 (rh'+h-1) G_{4,X}
-8 fA_0 \left[ rfh^2 A_1A_1'-(rf'h+rfh'+fh) X_1\right] G_{4,XX}
\notag\\
\hspace{-0.4cm}
& & 
-fA_0 \left[ f(3h-1)h'A_1+h(h-1) (f'A_1+2fA_1')   \right] G_{5,X}
-2 fhA_0X_1\left[2 fhA_1'+(f'h+fh')A_1 \right] G_{5,XX}
\notag\\
\hspace{-0.4cm}
& & 
-2 f \left[ f (3 h-1) h'A_0'+h (h-1) (2fA_0''-f'A_0') \right] G_6
-4 fh A_0'X_1 \left( h A_0 A_0'-2 fh^2 A_1 A_1'
-2 f'hX_0+2 fh'X_1 \right) G_{6,XX}
\notag\\
\hspace{-0.4cm}
& & 
-2 f \left[ 4 fh^2 X_1 A_0''-2 h (hX- X_0) f'A_0'
+2f (6h-1)h' X_1A_0'+h(h-1)A_0A_0'^2
-2 fh^2(3h-1) A_0'A_1A_1'\right] G_{6,X}
\notag\\
\hspace{-0.4cm}
& & 
-4 fh \left[ 2 rfh A_1 A_0''- \{(rf' h-3 rfh'-2fh) A_1-2 rfhA_1'\} A_0'  \right] g_5
\notag\\
\hspace{-0.4cm}
& & 
-4 rfh A_0' \left[ hA_0A_0'A_1+4 fhX_1A_1'-2 A_1(f'hX_0-fh'X_1)\right] g_{5,X}
=0\,.
\label{be4}
\ea
And similarly, the equation of motion for $A_1$ reads
\ba
\hspace{-0.4cm}
& & 
A_1 \left[ r^2fG_{2,X}-2 (rf'h+fh-f) G_{4,X}
+4h(rA_0 A_0'-rf' X-fX_1) G_{4,XX}
-hA_0'^2(3h-1) G_{6,X}-2h^2X_1 A_0'^2 G_{{6,{  XX}}}\right]
\nonumber \\
\hspace{-0.4cm}
& &=r [r(f' X-A_0 A_0')+4 fX_1] G_{3,X}
+2 f'hX_1G_{5,X}+(A_0 A_0'-f' X)\left[ (1-h)G_{5,X}-2 hX_1G_{5,XX}\right]
\nonumber \\
\hspace{-0.4cm}
& &
\hspace{.35cm}
-2rh A_0'^2( g_{5} +2 X_1 g_{5,X})\,.
\label{be5}
\ea
For the theories in which only the couplings $G_i(X)$ with 
even index $i$ are present, Eq.~(\ref{be5}) admits the branch 
of the solution $A_1=0$. This is not the case for the theories 
containing the couplings $G_{3}, G_{5}$ and $g_5$.
To derive the gravitational equations of motion, 
we write the metric (\ref{metric}) in a more general form 
$ds^2=-f(r) dt^{2} +h^{-1}(r) dr^{2}+ r^{2}e^{2\zeta(r)} 
(d\theta^{2}+\sin^{2}\theta\,d\varphi^{2})$ 
and express the action (\ref{action}) in terms of 
$f, h, \zeta, A_0, A_1$. 
Varying the resulting action with respect to $f, h, \zeta$ 
and setting $\zeta=0$ in the end, we obtain the following equations of motion 
\ba
\hspace{-0.7cm}
& &
\left( c_{1} + \frac{c_{2}}{r} + \frac{c_{3}}{r^{2}} \right) h' 
+ c_{4} + \frac{c_{5}}{r} + \frac{c_{6}}{r^{2}}
=0\,, 
\label{be1} \\
\hspace{-0.7cm}
& &-\frac{h}{f} \left( c_{1} + \frac{c_{2}}{r} + \frac{c_{3}}{r^{2}} \right) f' 
+ c_{7} + \frac{c_{8}}{r} + \frac{c_{9}}{r^{2}}
=0\,,
\label{be2}\\
\hspace{-0.7cm}
& & \left( c_{10} + \frac{c_{11}}{r} \right) f'' + \left( c_{12} + \frac{c_{13}}{r} \right) f'^{2}
+ \left( \frac{c_{2}}{2 f} + \frac{c_{14}}{r} \right) f' h' + \left( c_{15} + \frac{c_{16}}{r} \right) f'
+\left( -\frac{c_{8}}{2 h} + \frac{c_{17}}{r} \right) h' + c_{18} + \frac{c_{19}}{r}=0\,,
\label{be3}
\ea
where the coefficients $c_{1,2,\cdots,19}$ are given in Appendix. Among Eqs.~(\ref{be4})-(\ref{be3}) 
four of them are independent, so we will mostly employ 
Eqs.~(\ref{be4}), (\ref{be5}), (\ref{be1}), and (\ref{be2}) in the discussions below.

\subsection{RN solutions in GR}

As a warm up, we first review the solutions 
in GR characterized by the functions 
\be
G_4=\frac{M_{\rm pl}^2}{2}\,,\qquad 
G_2=G_3=G_5=G_6=0\,,\qquad g_5=0\,,
\ee
where $M_{\rm pl}$ is the reduced Planck mass. 
Then, Eqs.~(\ref{be4}), (\ref{be1}), and (\ref{be2}) 
reduce, respectively, to 
\ba
& &
2fh \left( r A_0''+2A_0' \right)
+r \left( fh'-hf' \right)A_0'=0\,,\label{be3GR}\\
& &
2f (rh'+h-1)M_{\rm pl}^2+r^2 h A_0'^2=0\,,\label{be1GR}\\
& &
2 \left[ rhf'+(h-1)f \right]M_{\rm pl}^2+r^2hA_0'^2=0\,,
\label{be2GR}
\ea
where Eq.~(\ref{be5}) is trivially satisfied. 
In this case, the longitudinal vector component corresponds to 
an unphysical gauge mode in that the value of $A_1$ 
is undetermined from Eqs.~(\ref{be3GR})-(\ref{be2GR}).
{}From Eqs.~(\ref{be1GR}) and (\ref{be2GR}) we have that 
$f'/f=h'/h$. The solution recovering the asymptotically flat geometry at spatial infinity $r \to \infty$ corresponds 
to $f=h$ after the proper rescaling of time.
Then, it follows that 
\ba
& & rA_0''+2A_0'=0\,,\label{A0ddeq}\\
& & f'=-\frac{r^2 A_0'^2+2M_{\rm pl}^2 (f-1)}
{2M_{\rm pl}^2r}\,,\label{fdeq}
\ea
which are integrated to give
\ba 
& & f=h=1-\frac{2M}{r}+\frac{Q^2}{2M_{\rm pl}^2r^2}\,,
\label{RNso}\\
& & A_0=P+\frac{Q}{r}\,,\label{RNso0}
\ea
where $P,Q,M$ are integration constants. 
This corresponds to the RN solution 
with mass $M$ and charge $Q$ of the BH. 
Note that $P$ is an arbitrary constant without having 
any physical meaning.

\subsection{Massive Proca field}

Let us also revisit the massive Proca field in GR given by the functions 
\ba
G_4=\frac{M_{\rm pl}^2}{2}\,,\qquad G_2=m^2 X\,,\qquad 
G_3=G_5=G_6=0\,,\qquad g_5=0\,,
\ea
where $m$ is a non-vanishing constant.
Since Eq.~(\ref{be5}) reduces to $A_1 r^2\,f\,m^2=0$, 
the longitudinal mode is constrained to be 
\be
A_1=0\,.
\ee
{}From Eqs.~(\ref{be4}), (\ref{be1}), and (\ref{be2}) it follows that 
\ba
& &
2fh \left( r A_0''+2A_0' \right)
+r \left( fh'-hf' \right)A_0'-2m^2frA_0=0\,,\label{be3Pro}\\
& &
2f (rh'+h-1)M_{\rm pl}^2+r^2 h A_0'^2+m^2r^2A_0^2=0\,,\label{be1Pro}\\
& &
2 \left[ rhf'+(h-1)f \right]M_{\rm pl}^2+r^2hA_0'^2
-m^2r^2A_0^2=0\,.
\label{be2Pro}
\ea
Combining Eq.~(\ref{be1Pro}) with Eq.~(\ref{be2Pro}), 
we obtain 
\be
\left( \frac{f}{h} \right)'=\frac{m^2 A_0^2\,r}{M_{\rm pl}^2h^2}\,.
\label{difffh}
\ee
On the horizon characterized by the distance $r_h$, 
we have that $f=h=0$. 
Since the metrics can be expanded as 
$f=\sum_{i=1}f_i(r-r_h)^i$ and $h=\sum_{i=1}h_i(r-r_h)^i$
around the horizon, the l.h.s. of Eq.~(\ref{difffh}) is finite 
at $r=r_h$. This means that $A_0$ needs to approach 0 as 
$r \to r_h$ for the consistency with the r.h.s. of 
Eq.~(\ref{difffh}). Imposing the asymptotic flatness at 
spatial infinity, $f\to 1$ and $h\to 1$ as $r\to\infty$,
it follows that $A_0 \to 0$ 
as $r \to \infty$. If we choose the boundary condition 
$A_0' \neq 0$ at $r=r_h$, then Eq.~(\ref{be3Pro}) gives 
rise to the growing-mode solution $A_0 \propto e^{mr}/r$ 
at spatial infinity. Provided that $A_0$ 
starts to deviate from 0 at some distance, 
this growing mode manifests for $r \gtrsim 1/m$.
Hence the solution consistent with the 
regularity in two asymptotic regimes is given by 
\be
A_0=0\,,
\label{A00}
\ee
throughout the horizon exterior \cite{Beken72}. 
Substituting Eq.~(\ref{A00}) into Eqs.~(\ref{be1Pro}) 
and (\ref{be2Pro}), the integrated solutions read
\be
f=h=1-\frac{2M}{r}\,,
\label{Schwa}
\ee
which corresponds to the Schwarzschild geometry.

The reason why we obtained the solution $A_0=0$ 
is attributed to the appearance of terms containing 
$A_0$ in Eqs.~(\ref{be3Pro})-(\ref{be2Pro}). 
Let us consider the more general case in which $G_2$ 
depends on both $X$ and $F$. 
{}From Eq.~(\ref{be5}) we have $A_1 r^2 f G_{2,X}=0$, 
so there exists the branch $A_1=0$ again for 
$G_{2,X} \neq 0$.
For the massive Proca field discussed above, 
the terms $-2r^2f^2A_0G_{2,X}$ in Eq.~(\ref{be4}), 
$G_2-2X_0G_{2,X}$ in Eq.~(\ref{be1}), and 
$-G_2+2X_1G_{2,X}$ in Eq.~(\ref{be2}) give rise to 
those containing $A_0$ in 
Eqs.~(\ref{be3Pro})-(\ref{be2Pro}). 
For the theories with $G_2=g_4(X)h(F)$, where 
$g_4(X)$ and $h(F)$ are functions of $X$ and $F$ 
respectively, they are factored out by $h(F)$. 
Since $F=hA_0'^2/(2f)$, the terms containing 
$A_0$ can be multiplied by the power of the 
derivative $A_0'$.
Then, the equations corresponding to (\ref{be3Pro}) 
and (\ref{difffh}) admit the solution where $A_0$ 
approaches a non-vanishing constant $P$ at spatial infinity 
[like the solution (\ref{RNso0})].

The above argument shows that the property (\ref{A00}) 
does not generally hold for the theories with $G_2=g_4(X)h(F)$.
As we will discuss later in Sec.~\ref{g45powersec}, 
this is actually the case for the coupling $G_2=-2g_4(X)F$.
If we consider the theories with $G_2=g(X)+h(F)$, the terms 
$A_0$ appear in the equations of motion without the 
multiplication of the powers of $A_0'$. In such cases, 
$A_0$ is generally forced to vanish. 

The coupling $G_2(X)$ is a rather specific case in which 
a non-vanishing effective mass term 
$G_{2,X}A_0$ is present in Eq.~(\ref{be4}) 
even at spatial infinity.
This fact does not allow the existence of hairy BH solutions.
For other derivative interactions $G_{3,4,5,6}$ and $g_5$
the terms containing $A_0$ are typically multiplied by the powers of $A_0'$ or by the inverse powers of $r$, so 
there exists the solution whose asymptotic 
behavior for $r \to \infty$ is $A_0 \to P \neq 0$.
In such cases, it is possible to realize hairy BH solutions 
with $A_0 \neq 0$ outside the horizon.
We also note that, in the presence of general derivative interactions, there are branches of solutions where 
the longitudinal mode $A_1$ does not vanish.

\section{Exact BH solutions for the quartic coupling $G_4$}
\label{exactsec}

In this section, we first revisit the exact hairy BH solution 
with $A_1 \neq 0$ which are known to exist for the theory given by the coupling $G_4(X)=M_{\rm pl}^2/2+X/4$ \cite{Chagoya} and then derive another solution 
corresponding to the branch $A_1=0$.
The exact BH solutions of Ref.~\cite{Chagoya} obeys 
the two conditions
\ba
&& f=h\,,\label{exactcon1}\\
&& X=X_c\,,
\label{exactcon2}
\ea
where $X_c$ is a constant. 
These two conditions are imposed to search for 
exact solutions in this section and Sec.~\ref{exactsec2}.
The condition (\ref{exactcon2}) translates to 
\be
A_1=\epsilon \frac{\sqrt{A_0^2-2fX_c}}{f}\,,
\label{A1A0re}
\ee
where $\epsilon=\pm 1$, and we used Eq.~(\ref{exactcon1}).

The longitudinal mode (\ref{A1A0re}) exhibits the divergence 
at the horizon where $f=0$ for $A_0 \neq 0$.
However, this divergence simply comes from the choice of 
coordinates. To see this, we introduce the tortoise coordinate 
$dr_* \equiv dr/f(r)$ and consider the scalar product \cite{Minami}
\be
A_{\mu}dx^{\mu}=A_0(r)dt+A_1(r)dr\,.
\label{product}
\ee
Since $A_1 \simeq \epsilon A_0/f$ around the horizon,  
the product (\ref{product}) reduces to 
\be
A_{\mu}dx^{\mu} \simeq A_0(r) 
\left( dt \pm dr_* \right)
=A_0(r) du_{\pm}\,,
\label{product2}
\ee
where $u_{+} \equiv t+r_*$ and 
$u_{-} \equiv t-r_*$ are the advanced and 
retarded null coordinates, respectively. 
The coordinates $u_{+}$ and $u_{-}$ are regular at the 
future and past event horizons, respectively.
Thus, the regularity of the vector field is ensured at the 
corresponding (future or past) horizon.

For the general quartic coupling $G_4(X)$ 
the vector-field Eqs.~(\ref{be4}) and (\ref{be5}) 
reduce, respectively, to 
\ba
\hspace{-0.6cm}
&& 
A_0''+\frac{2}{r}A_0'-\frac{A_0'}{2} 
\left( \frac{f'}{f}- \frac{h'}{h} \right)
+\frac{2A_0G_{4,X}}{hr^2} 
\left( rh'+h-1 \right)
-\frac{2A_0A_1 G_{4,XX}}{fr^2}
\left[ 2fh rA_1'+(fh+fh'r+hf'r)A_1 \right]=0\,,
\label{be4G4} \\
\hspace{-0.6cm}
&&
A_1 \left[ f \{ hrf'+ (h-1)f\}G_{4,X}
-h (A_1^2 ff'h r+A_1^2 f^2 h-A_0^2 f'r
+2A_0A_0' fr)G_{4,XX} \right]=0\,.
\label{be4G5}
\ea
{}From Eq.~(\ref{be4G5}) there are two branches 
characterized by $A_1 \neq 0$ and $A_1=0$.
In the following, we will consider the two cases separately.

\subsection{$A_1 \neq 0$}

If the second derivative of $G_4$ with respect to $X$ 
obeys the condition
\be
G_{4,XX}(X_c)=0\,,
\label{G4XX}
\ee
then Eq.~(\ref{be4G5}) can be satisfied for 
\be
hrf'+ (h-1)f=0\,,
\label{hfeq}
\ee
with $G_{4,X} \neq 0$. 
Under the condition (\ref{exactcon1}), the solution 
to Eq.~(\ref{hfeq}) is given by the Schwarzschild metric
\be
f=h=1-\frac{2M}{r}\,,
\label{fhso}
\ee
where $M$ is an integration constant.
Then, Eq.~(\ref{be4G4}) is satisfied for 
\be
A_0''+\frac{2}{r}A_0'=0\,,
\ee
whose integrated solution is 
\be
A_0=P+\frac{Q}{r}\,,
\label{A0so}
\ee
where $P$ and $Q$ are constants. 
Now, we search for solutions obeying the condition (\ref{exactcon2}). 
On using Eqs.~(\ref{A1A0re}), (\ref{fhso}), (\ref{A0so}), 
and their derivatives with respect to $r$, 
we find that Eqs.~(\ref{be1})-(\ref{be3}) can be satisfied for 
\ba
& &G_{4,X}(X_c)=\frac{1}{4}\,,\label{G4X}\\
& &X_c=\frac{P^2}{2}\,,
\ea
with the longitudinal vector component
\be
A_1=\epsilon \frac{\sqrt{2P(MP+Q)r+Q^2}}
{r-2M}\,.
\label{A1ep}
\ee
Since the constant $P$ in Eq.~(\ref{A0so}) does not depend on 
$M$ and $Q$, the Proca hair is of the primary 
type \cite{Herdeiro}.
The function $G_4(X)$ obeying 
the two conditions (\ref{G4XX}) and (\ref{G4X}) is given by  
\be
G_4(X)=G_4(X_c)+\frac{1}{4} (X-X_c) 
+\sum_{n=3} b_n(X-X_c)^n\,,
\label{G4exact}
\ee
where $X_c=P^2/2$, and $b_n$'s are constants.
The model $G_4(X)=M_{\rm pl}^2/2+X/4$ 
of Ref.~\cite{Chagoya} is the special case of Eq.~(\ref{G4exact}), 
i.e., $G_4(X_c)=M_{\rm pl}^2/2+X_c/4$ and $b_n=0$ 
for $n \geq 3$. 
The above solution is a stealth Schwarzschild solution with 
a non-vanishing longitudinal vector component.

\subsection{$A_1=0$}

Let us proceed to another branch characterized by $A_1=0$. 
Imposing the condition (\ref{exactcon2}), we have 
$A_0^2(r)=2f(r)X_c$ from Eq.~(\ref{A1A0re}). 
Under the condition (\ref{exactcon1}), 
Eq.~(\ref{be4}) reduces to 
\be
r \left( 2r f f''-rf'^2+4ff' \right)
+8f \left( rf'+f-1 \right)G_{4,X}=0\,.
\label{G4eq4}
\ee
If we consider the case in which the relation  $rf'+f-1=0$ 
holds, then the resulting solution $f=1-2M/r$ does not 
obey Eq.~(\ref{G4eq4}). 
Then, we search for solutions satisfying 
\be
2r f f''-rf'^2+4ff'=0\,,
\label{ddf}
\ee
with 
\be
G_{4,X}(X_c)=0\,.
\label{G4Xcon}
\ee
Integration of Eq.~(\ref{ddf}) leads to
\be
f=\left( C-\frac{M}{r} \right)^2\,,
\label{fex}
\ee
where $C$ and $M$ are constants.
The solution (\ref{fex}) is consistent with  
Eqs.~(\ref{be1})-(\ref{be3}) for $C=1$ and
\be
G_4(X_c)=\frac{X_c}{2}\,.
\label{G4con}
\ee 
An explicit model satisfying the conditions (\ref{G4Xcon}) 
and (\ref{G4con}) is given by 
\be
G_4(X)=\frac{X_c}{2}
+\sum_{n=2} b_n\left( X-X_c \right)^n\,.
\label{G4A10}
\ee
The resulting exact solution reads
\be
f=h=\left( 1-\frac{M}{r} \right)^2\,,\qquad
A_0=P-\frac{MP}{r}\,,\qquad 
A_1=0\,,
\label{extremal}
\ee
where $P=\epsilon \sqrt{2X_c}$.
This corresponds to the extremal RN BH solution.  

\section{Exact BH solutions for general couplings}
\label{exactsec2}

We proceed to the derivation of exact BH solutions in the presence of the couplings $G_3(X), G_5(X), G_6(X), 
g_5(X)$ and $G_2(X,F)=-2g_4(X)F$.
Throughout the analysis we take into account the 
Einstein-Hilbert term $M_{\rm pl}^2/2$ in $G_4(X)$.
Analogous to the derivation of exact solutions given in 
Sec.~\ref{exactsec}, we will impose the two conditions (\ref{exactcon1}) and (\ref{exactcon2}) in the 
following discussion.

\subsection{Cubic coupling $G_3(X)$}

For the cubic interaction $G_3(X)$, Eq.~(\ref{be5}) reduces to 
\be
G_{3,X} \left[ f^2(rf'+4f) A_1^2
+r(2f A_0'-f'A_0)A_0 \right]=0\,.
\label{G3Xeq}
\ee
Since there are two branches satisfying
(i) $G_{3,X}(X_c)=0$ and (ii) $G_{3,X}(X_c) \neq 0$,  
we will discuss each case separately.

\subsubsection{$G_{3,X}(X_c)=0$}
\label{G3subsec}

For the branch (i), Eqs.~(\ref{be4}) and (\ref{be1}) reduce 
to Eqs.~(\ref{A0ddeq}) and (\ref{fdeq}), respectively, 
so we obtain the RN solutions (\ref{RNso}) and (\ref{RNso0}).
{}From Eq.~(\ref{A1A0re}) the longitudinal mode 
reduces to 
\be
A_1=\epsilon \frac{2M_{\rm pl}r 
\sqrt{M_{\rm pl}^2 (P^2-2X_c)r^2
+2M_{\rm pl}^2(PQ+2MX_c)r+Q^2(M_{\rm pl}^2-X_c)}}
{2M_{\rm pl}^2 (r^2-2Mr)+Q^2}\,.
\label{G3A1so}
\ee
Since the constant $P$ is independent of $M$ and $Q$, 
it can be regarded as the primary hair.
A concrete example realizing this exact solution is given by 
\be
G_3(X)=G_3(X_c)+\sum_{n=2} b_n\left( X-X_c \right)^n\,.
\label{G3exactlag}
\ee
\subsubsection{$G_{3,X}(X_c) \neq 0$}

The branch (ii) includes  the case of vector Galileons ($G_3=\beta_3 X$). On using the conditions (\ref{exactcon1}) and (\ref{exactcon2}) 
in Eq.~(\ref{G3Xeq}), it follows that 
\be
f'=\frac{rA_0A_0'+2A_0^2-4fX_c}{X_c r}\,.
\label{dfeq}
\ee
Substituting this relation into Eq.~(\ref{be4}), we obtain $rA_0''+2A_0'=0$. Hence the integrated solution is 
$A_0=P+Q/r$ with two constants $P$ and $Q$.
Then, Eq.~(\ref{dfeq}) is integrated to give
\be
f=\frac{1}{2X_c} \left( P+\frac{Q}{r} \right)^2+\frac{C}{r^4}\,,
\ee
where $C$ is a constant.  To satisfy the asymptotically flat boundary condition $f \to 1$ as $r \to \infty$, we require that $P^2=2X_c$. The above solutions are consistent with 
Eqs.~(\ref{be1})-(\ref{be3}) for $C=0$ and 
$X_c=M_{\rm pl}^2$. 
On defining $M=\pm Q/(\sqrt{2}M_{\rm pl})$ for 
$P=\mp \sqrt{2}M_{\rm pl}$, we obtain the extremal 
BH solution (\ref{extremal}) with 
$P=\epsilon \sqrt{2}M_{\rm pl}$.
The longitudinal mode $A_1$ vanishes for this exact solution.

\subsection{Quintic coupling $G_5(X)$}

For the quintic interaction $G_5(X)$, combining Eq.~(\ref{be4}) with Eq.~(\ref{be5}) leads to
\ba
&& 
r A_0''+2A_0'=0\,,\label{be4d} \\
&&
\left( A_0^2-2fX_c \right) \left( A_0A_0'-X_c f' \right)
G_{5,XX}(X_c)-\left[A_0^2 f'+A_0A_0'(f-1)
-(3f-1)f'X_c \right]G_{5,X}(X_c)=0\,.
\label{be5d}
\ea
The solution to Eq.~(\ref{be4d}) is given by 
$A_0=P+Q/r$. If
\be
G_{5,X}(X_c)=0\,,
\label{G5Xc}
\ee
then Eq.~(\ref{be5d}) is satisfied either for 
(i) $A_0A_0'=X_cf'$ or (ii) $A_0^2=2fX_c$.

For the branch (i) we have
\be
f'=\frac{A_0A_0'}{X_c}
=-\frac{(Pr+Q)Q}{X_cr^3}\,,
\ee
which is integrated to give
$f=C-2M/r+Q^2/(2X_c r^2)$ 
with $M=-PQ/(2 M_{\rm pl}^2)$. 
For the consistency with Eqs.~(\ref{be1})-(\ref{be2}) 
we require that $C=1$ and $X_c=M_{\rm pl}^2$, 
so we obtain the RN solution 
\be
f=h=1-\frac{2M}{r}+\frac{Q^2}{2M_{\rm pl}^2r^2}\,,
\label{RNsolu}
\ee
with the vector components
\be
A_0=-\frac{2MM_{\rm pl}^2}{Q}+\frac{Q}{r}\,,\qquad
A_1=\epsilon \frac{2M_{\rm pl}^3 
\sqrt{2(2M^2M_{\rm pl}^2-Q^2)}\,r^2}
{Q[2M_{\rm pl}^2r(2M-r)-Q^2]}\,.
\label{A01G5}
\ee
The existence of this solution requires the condition 
$2M^2M_{\rm pl}^2>Q^2$. 
Since the Proca hair $P=-2MM_{\rm pl}^2/Q$ 
is fixed by $M$ and $Q$, it is of the secondary type.

{}From Eq.~(\ref{A1A0re})
the branch (ii) corresponds to $A_1=0$. In this case, the integration of Eqs.~(\ref{be1})-(\ref{be2}) gives rise to the RN solutions (\ref{RNso}) and (\ref{RNso0}).
On using the property $A_0^2=2fX_c$, the metric $f$ reduces to the extremal RN solution $f=(1-M/r)^2$ with 
the particular relation $Q^2=2M^2M_{\rm pl}^2$. 
Indeed, this case can be regarded as the special case of 
the solutions (\ref{A01G5}) with $A_1=0$.

A concrete mode realizing the above  
solutions is given by 
\be
G_5(X)=G_5(X_c)+\sum_{n=2} b_n 
\left( X-X_c \right)^n\,,
\label{G5explicit}
\ee
where $X_c=M_{\rm pl}^2$.

\subsection{Sixth-order coupling $G_6(X)$}

In the presence of the sixth-order coupling $G_6(X)$, 
Eq.~(\ref{be5}) reduces to
\be
A_0'^2 A_1 \left[ A_1^2h^2 G_{6,XX}
+\left(1-3h\right)G_{6,X}  \right]=0\,.
\label{longG6}
\ee
Let us search for exact solutions satisfying either $A_0'=0$ or $A_1=0$.

\subsubsection{$A_0'=0$}
\label{G6A0v}

In this case we have
\be
A_0=P={\rm constant}\,,
\ee
under which Eq.~(\ref{be4}) is trivially satisfied. 
{}From Eqs.~(\ref{be1}) and (\ref{be2}) we obtain 
$rf'+f-1=0$, so the integrated solution is given by 
the Schwarzschild metric
\be
f=h=1-\frac{2M}{r}\,.
\label{fhG6}
\ee
In fact, this solution exists for general couplings $G_6(X)$ 
with any value of $A_1$. 
In the present case the longitudinal mode is subject to 
the constraint (\ref{A1A0re}), so it is given by 
\be
A_1=\epsilon \frac{\sqrt{r(
P^2r+4MX_c-2rX_c)}}
{r-2M}\,.
\label{A1ex}
\ee
Since $A_1$ approaches the constant $\epsilon \sqrt{P^2
-2X_c}$ as $r \to \infty$, we require the 
condition $
P^2>2X_c$ for the existence of this solution. 

\subsubsection{$A_1=0$}

We proceed to the case in which the longitudinal 
mode obeys
\be
A_1=0\,.
\label{A1G6}
\ee
In this case we have $A_0^2(r)=2f(r)X_c$, so
we take the $r$-derivative of this relation and substitute 
them into Eqs.~(\ref{be4})-(\ref{be1}). 
Then, Eqs.~(\ref{be3}) and (\ref{be4}) reduce, respectively, to 
\ba
& &
2M_{\rm pl}^2 f \left( rf''+2f' \right)
-X_cr f'^2-4X_cff'f''G_6=0\,,\label{G6so1}\\
& &
2r f\left( rf''+2f' \right)-r^2f'^2-2\left[2f^2 f''+f(f'^2-2f'')+f'^2 
\right]G_6 +2X_c (f-1)f'^2G_{6,X}=0\,.\label{G6so2}
\ea
Let us search for exact solutions satisfying the two conditions
\be
G_{6}(X_c)=0\,,\qquad G_{6,X}(X_c)=0\,.
\label{G6con}
\ee
{}From Eqs.~(\ref{G6so1})-(\ref{G6so2}) we obtain the 
integrated solution $f=(C-M/r)^2$ with $X_c=M_{\rm pl}^2$. 
Since the integration constant is fixed to be $C=1$ from 
Eq.~(\ref{be1}), we obtain the extremal RN solution 
(\ref{extremal}) with $P=\epsilon \sqrt{2}M_{\rm pl}$. 
This is equivalent to the solution derived for the quintic 
coupling $G_5(X)$ with the branch $A_0^2=2fX_c$.

A concrete model realizing this solution is 
\be
G_6(X)=\sum_{n=2}b_n \left( X-X_c \right)^n\,,
\label{G6explicit}
\ee
where $X_c=M_{\rm pl}^2$.

\subsection{Quartic intrinsic vector-mode coupling $g_4(X)$}

Let us consider the coupling given by 
\be
\label{G2g4}
G_2(X,F)=-2g_4(X)F\,,
\ee
where $g_4(X)$ is a function of $X$. 
This corresponds to the intrinsic vector mode originally 
introduced in ${\cal L}_4$ as a form 
$g_4(X)(\nabla_{\rho}A_{\sigma}\nabla^{\rho}A^{\sigma}
-\nabla_{\rho}A_{\sigma}\nabla^{\sigma}A^{\rho})$ 
with $g_4(X)=c_2G_{4,X}$ \cite{Heisenberg}.
Then, Eq.~(\ref{be5}) reduces to 
\be
g_{4,X} A_0'^2 A_1=0\,.
\label{A02A1}
\ee
Let us consider the case in which the relation  
\be
g_{4,X}(X_c)=0
\ee
is satisfied. 
{}From Eqs.~(\ref{be4}) and (\ref{be1}) it follows that 
\ba
& &
\left( 2g_4-1 \right) \left( r A_0''+2A_0' \right)=0\,,\\
& &
2 \left( rf'+f-1 \right) M_{\rm pl}^2-(2g_4-1)r^2A_0'^2=0\,.
\label{g5eq1}
\ea
For $g_4 (X_c) \neq 1/2$ these equations are 
integrated to give 
\ba
& &
A_0=P+\frac{Q}{r}\,,\\
& & 
f=h=1-\frac{2M}{r}+\frac{Q^2}{2M_{\rm pl}^2 r^2} 
\left[1-2g_4(X_c) \right]\,,
\ea
with the longitudinal mode (\ref{A1A0re}).
The metric is of the RN type with the effective charge
$Q_{\rm eff}= \sqrt{1-2g_4(X_c) } Q$, 
which is different from $Q$ unless $g_4(X_c)=0$.

A concrete model realizing this solution is given by 
\be
g_4(X)=g_4(X_c)+\sum_{n=2}b_n \left( X-X_c \right)^n\,.
\label{g4exact}
\ee
If $g_4(X_c)=1/2$, then the Schwarzschild solution 
$f=h=1-2M/r$ follows from Eq.~(\ref{g5eq1}) 
with $A_0$ undetermined.
This comes from the fact that, for $g_4(X_c)=1/2$, 
the Lagrangian $F$ is compensated 
by the term $-2g_4F$.

It is also possible to satisfy Eq.~(\ref{A02A1}) either for 
(i) $A_0'=0$ or (ii) $A_1=0$.
For the branch (i), we obtain the Schwarzschild solution 
$f=h=1-2M/r$ with $A_0={\rm constant}$ and $A_1$ 
given by Eq.~(\ref{A1A0re}) for general couplings $g_4(X)$.
For the branch (ii),  there exists an exact solution under 
the conditions $g_{4,X}(X_c)=0$ and $g_{4}(X_c)=1/2$. 
In this case, the resulting solution reads 
\be
f=h=1-\frac{2M}{r}\,,\qquad 
A_0=\epsilon \sqrt{2 \left( 1-\frac{2M}{r}\right)X_c}\,,
\qquad A_1=0\,.
\ee
This solution exists for the function (\ref{g4exact}) with 
$g_{4}(X_c)=1/2$.

\subsection{Quintic intrinsic vector-mode coupling $g_5(X)$}

Let us finally proceed to the exact solution for the quintic coupling $g_5(X)$.
Then, Eq.~(\ref{be5}) reduces to 
\be
A_0'^2 \left[ fg_5-\left( A_0^2-2 f X_c \right)g_{5,X} 
\right]=0\,.
\ee

For the branch $A_0'=0$, the Schwarzschild solution 
$f=h=1-2M/r$ follows with $A_1$ given by Eq.~(\ref{A1A0re})
for general couplings $g_5(X)$.

For the other branch $f g_5=( A_0^2-2 f X_c)g_{5,X}$, there exists 
an exact solution under the condition 
\be
g_{5,X}(X_c)=0\,.
\ee
{}From Eqs.~(\ref{be4}) and (\ref{be1}) we obtain the equations 
same as Eqs.~(\ref{A0ddeq}) and (\ref{fdeq}), respectively, 
so the integrated solutions to $f,h,A_0$ yield the RN solutions 
(\ref{RNso}) and (\ref{RNso0}) with $A_1$ given 
by Eq.~(\ref{A1A0re}). 
Since $g_5(X_c)=0$ in this case, the quintic interaction 
in the form 
\be
g_5(X)=\sum_{n=2}b_n \left( X-X_c \right)^n
\label{g5exact}
\ee
gives rise to the RN solution with the non-vanishing 
longitudinal mode.

\section{Power-law cubic couplings $G_3(X)$}
\label{G3powersec}

In Secs.~\ref{exactsec} and \ref{exactsec2}, we have imposed the two 
conditions (\ref{exactcon1}) and (\ref{exactcon2}) for the purpose 
of deriving exact BH solutions. 
We now focus on the solutions where $f \neq h$ and $X$ is not constant. 
Numerical works are generally required to find such non-exact solutions.
In this section, we first study the model in which the function $G_{3}$ 
is given by the power-law function $X^n$, where $n$ is assumed 
to be a positive integer. 
In the subsequent sections, we will study the models in which 
the functions $G_4,G_5,G_6, g_4,g_5$ contain the function $X^n$. 
In the whole analysis by the end of Sec.~\ref{g45powersec}, 
we include the Einstein-Hilbert term $M_{\rm pl}^2/2$ in $G_4$.
We will focus on the asymptotically flat solutions and 
not take into account the vector-field mass and 
the cosmological constant.

We begin with the power-law cubic coupling model
given by 
\be
G_3=\beta_3 M_{\rm pl}^2 \left( 
\frac{X}{M_{\rm pl}^2} \right)^n\,,
\label{G3powermodel}
\ee
with $G_4=M_{\rm pl}^2/2$, where
$\beta_3$ is a dimensionless constant.
In the following, we will discuss the cases of 
$n=1$ (vector Galileons) and $n \geq 2$, separately.

\subsection{$n=1$}

{}From (\ref{be5}) the longitudinal component for $n=1$ 
is related to $A_0,f,h$, as
\be
A_1=\epsilon \sqrt{\frac{rA_0 (f'A_0-2fA_0')}
{fh (rf'+4f)}}\,.
\label{A1be3}
\ee
We substitute Eq.~(\ref{A1be3}) and the $r$-derivative of it
into Eqs.~(\ref{be4}), (\ref{be1}), and (\ref{be2}) 
to eliminate the $A_1$ dependence.

Around the BH horizon characterized by the distance $r_h$, 
we expand $f,h,A_0$ in the following forms
\be
f=\sum_{i=1}^{\infty} f_i(r-r_h)^i\,,\qquad
h=\sum_{i=1}^{\infty} h_i(r-r_h)^i\,,\qquad
A_0=a_0+\sum_{i=1}^{\infty} a_i(r-r_h)^i\,,
\label{fhA0}
\ee
where $f_i,h_i,a_0,a_i$ are constants. 
The effect of the coupling $\beta_3$ works as corrections 
to the RN metrics given by 
\be
f_{\rm RN}=h_{\rm RN}
=\left( 1-\frac{r_h}{r} \right)
\left( 1- \mu \frac{r_h}{r} \right)\,.
\label{fRN}
\ee
The constant $\mu$ is in the range $0<\mu<1$, 
so that $r_h$ corresponds to the outer horizon.
Compared to Eq.~(\ref{RNso}), there is the correspondence 
$Q^2=2r_h(2M-r_h)M_{\rm pl}^2$ with the inner horizon 
$\tilde{r}_h=2M-r_h$. Hence the constant $\mu$ is given by 
$\mu=2M/r_h-1$ with $M<r_h<2M$. 
To derive the coefficients $f_i,h_i,a_0,a_i$ in Eq.~(\ref{fhA0})
iteratively, we assume that $f_1,h_1,a_0$ are positive 
and choose the positive branch of Eq.~(\ref{A1be3}) 
for $r>r_h$. We also take the contributions up to 
linear order in $\beta_3$ under the assumption that 
the coupling $\beta_3$ works as a correction 
to the RN solutions. Up to 
the order of $(r-r_h)^2$, the coefficients are given by 
\be
f_1=h_1=\frac{1-\mu}{r_h}\,,\qquad 
a_1= \frac{\sqrt{2\mu}M_{\rm pl}}{r_h}\,,
\label{f1a1}
\ee
and
\be
f_2 = \frac{2\mu-1}{r_h^2}
+{\cal F}_2 
\beta_3\,,\qquad
h_2 = \frac{2\mu-1}{r_h^2}+{\cal H}_2 
\beta_3\,,\qquad
a_2 = -\frac{\sqrt{2\mu}M_{\rm pl}}{r_h^2}+\alpha_2 \beta_3\,,
\label{h2}
\ee
where 
\ba
{\cal F}_2 &=&
\frac{1-\mu}{M_{\rm pl}}\frac{\sqrt{2\mu}M_{\rm pl}+4a_0}
{2\sqrt{2\mu}a_0+(1+\mu)M_{\rm pl}}\alpha_2
\,,\qquad
{\cal H}_2 =
-\frac{1-\mu}{M_{\rm pl}}\frac{3\sqrt{2\mu}M_{\rm pl}+4a_0}
{2\sqrt{2\mu}a_0+(1+\mu)M_{\rm pl}}\alpha_2\,, 
\nonumber \\
\label{calH2}
\alpha_2 &=& 
-\frac{\left[(1+\mu) M_{\rm pl}+2\sqrt{2\mu} a_0\right]
\left[\mu M_{\rm pl}\left\{(1-\mu)M_{\rm pl}-\sqrt{2\mu}a_0\right\}
-(3-\mu)a_0^2\right]}{(1-\mu)^3 M_{\rm pl} r_h}
\,.\label{al2}
\ea
{}From Eq.~\eqref{f1a1} and the condition $0<\mu<1$, 
the quantity $h_1r_h$ is in the range $0<h_1r_h<1$. 
The corrections to the RN solutions from 
the coupling $\beta_3$ arise at the order of $(r-r_h)^2$ for $f,h,A_0$. 

Taking the positive branch of
Eq.~(\ref{A1be3}), the behavior of 
the longitudinal mode around the horizon is given by 
\be
A_1=\frac{a_0}{f_1(r-r_h)}
-\frac{a_0 [(f_2+h_2)r_h+4f_1]}
{2f_1^2r_h}+{\cal O}(r-r_h)\,,
\label{A1ho}
\ee
which exhibits the divergence at $r=r_h$. 
Analogous to the discussion given after
Eq.~(\ref{A1A0re}), the scalar product (\ref{product}) 
reduces to $A_\mu dx^{\mu} \simeq a_0 du_{+}$ around 
$r=r_h$ for the solution (\ref{A1ho}).
Hence the regularity of the vector field is ensured at the future horizon.

We also derive asymptotic flat solutions 
satisfying $f,h \to 1$, and $A_0 \to P$
as $r \to \infty$, where $P$ is a constant. 
To obtain the solutions at spatial infinity, we expand 
$f,h,A_0$ as the power series of $1/r$, as
\be
f=1+\sum_{i=1}^{\infty}\frac{\tilde{f}_i}{r^i}\,,\qquad
h=1+\sum_{i=1}^{\infty}\frac{\tilde{h}_i}{r^i}\,,\qquad
A_0=P+\sum_{i=1}^{\infty}\frac{\tilde{a}_i}{r^i}\,.
\label{fh}
\ee
For the cubic coupling model (\ref{G3powermodel}) 
with $n=1$, there exists an asymptotic solution 
where the longitudinal mode is given by 
$A_1=\sum_{i=1}^{\infty}\tilde{b}_i/r^i$. 
Substituting this expression of $A_1$ and Eqs.~(\ref{fh})
into Eqs.~(\ref{be4})-(\ref{be3}), we obtain the following 
iterative solutions 
\ba
f&=& 1-\frac{2M}{r}-\frac{P^2M^3}{6M_{\rm pl}^2 r^3}
+\frac{M^4P^2(P^2-2M_{\rm pl}^2)+3M_{\rm pl}^2
\tilde{b}_2^2}
{3M_{\rm pl}^2(2M_{\rm pl}^2-P^2)r^4} \nonumber \\
& &-\frac{M[3\beta_3M^4P^4 (P^2+14M_{\rm pl}^2)+
16M_{\rm pl}^4 (8\tilde{b}_2M+3\beta_3 \tilde{b}_2^2
-6\beta_3M^4 P^2)]}{80\beta_3 M_{\rm pl}^4 (P^2-2M_{\rm pl}^2)r^5}
+{\cal O} \left( 
\frac{1}{r^6} \right)\,,\label{G3f} \\
h&=& 1-\frac{2M}{r}-\frac{P^2M^2}{2M_{\rm pl}^2 r^2}
-\frac{P^2M^3}{2M_{\rm pl}^2 r^3}
+\frac{2M^4P^2(P^2-2M_{\rm pl}^2)+12M_{\rm pl}^2\tilde{b}_2^2}
{3M_{\rm pl}^2(2M_{\rm pl}^2-P^2)r^4} \nonumber \\
& &
-\frac{M[\beta_3M^4P^4 (P^2+46M_{\rm pl}^2)+
48M_{\rm pl}^4 (8\tilde{b}_2M+\beta_3 \tilde{b}_2^2
-2\beta_3M^4 P^2)]}
{48\beta_3 M_{\rm pl}^4(P^2-2M_{\rm pl}^2)r^5}
+{\cal O} \left( 
\frac{1}{r^6} \right)\,,\label{G3h} \\
A_0&=& 
P-\frac{PM}{r}-\frac{PM^2}{2r^2}
-\frac{PM^3(P^2+6M_{\rm pl}^2)}
{12M_{\rm pl}^2r^3}
-\frac{P^2M^4(2P^2+5M_{\rm pl}^2)
(P^2-2M_{\rm pl}^2)+8M_{\rm pl}^4\tilde{b}_2^2}
{8PM_{\rm pl}^2(P^2-2M_{\rm pl}^2)r^4} \nonumber \\
& &
-\frac{M}{480r^5}\left[\frac{M^4P(P^2+30M_{\rm pl}^2)
(9P^2+14M_{\rm pl}^2)}{M_{\rm pl}^4}
+\frac{48\tilde{b}_2\{\beta_3\tilde{b}_2(3P^2+10M_{\rm pl}^2)
+16MM_{\rm pl}^2\}}{\beta_3 P (P^2-2 M_{\rm pl}^2)}\right]
+{\cal O} \left( \frac{1}{r^6} \right)\,,\label{G3A0} \\
A_1&=&
\frac{\tilde{b}_2}{r^2}+
\frac{M(M+2\tilde{b}_2 \beta_3)}{\beta_3 r^3}
+\frac{12M^3M_{\rm pl}^2+\tilde{b}_2M^2 (P^2+16M_{\rm pl}^2)
\beta_3}{4\beta_3 M_{\rm pl}^2 r^4} \nonumber \\
& & +\frac{1}{3M_{\rm pl}^2r^5} \left[ 
\frac{M^4(P^2+22M_{\rm pl}^2)}{\beta_3}
+4\tilde{b}_2 \left\{ M^3 (P^2+6M_{\rm pl}^2)
+\frac{3\tilde{b}_2M_{\rm pl}^4}{\beta_3P^2(P^2-2M_{\rm pl}^2)} 
\right\} \right]
+{\cal O} \left( 
\frac{1}{r^6} \right)\,,
\label{G3A1}
\ea
where we have set $\tilde{f}_1=\tilde{h}_1=-2M$. 
The metric $f$ does not contain the term proportional to 
$P^2M^2/(M_{\rm pl}^2r^2)$ unlike the metric $h$, 
so there is the difference between $f$ and $h$ for $P \neq 0$ 
at the order of $1/r^2$. The leading-order solutions to 
temporal and longitudinal vector components are given, 
respectively, by $A_0 \simeq P(1-M/r)$ and 
$A_1 \simeq \tilde{b}_2/r^2$, which are also consistent 
with the solutions derived under the weak gravity approximation far outside a spherically symmetric 
body \cite{DeFeliceVain}.
The effects of the constants $\tilde{b}_2$ and $\beta_3$ start to appear in the metrics $f$ and $h$ 
at the orders of $1/r^4$ and $1/r^5$, respectively.

\begin{figure}
\begin{center}
\includegraphics[height=3.5in,width=3.5in]{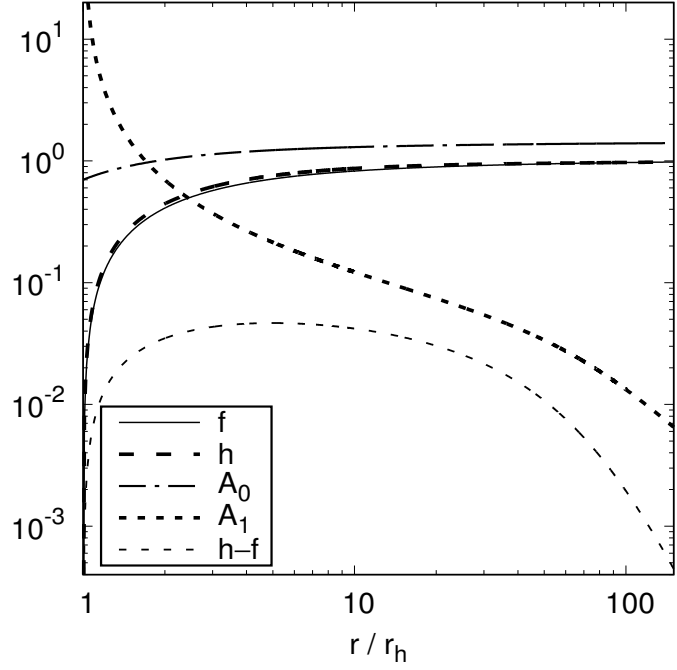}
\end{center}
\caption{\label{fig1}
Numerical solutions of $f, h,A_0, A_1,h-f$ outside the horizon 
for the cubic vector Galileon model $G_3(X)=\beta_3X$ 
with the coupling $\beta_3=7\times10^{-3}/(r_h M_{\rm pl})$.
The boundary conditions around $r=r_h$ are chosen to satisfy Eqs.~(\ref{f1a1}) and (\ref{h2}) with 
$\mu=0.2$, $a_0=0.7M_{\rm pl}$ 
at $r=1.001r_h$. The two asymptotic solutions in the regimes 
$r \simeq r_h$ and $r \gg r_h$ smoothly join each other.}
\end{figure}

To study whether the near-horizon solutions (\ref{fhA0}) 
connect to the large-distance solutions (\ref{fh}), 
we numerically solve Eqs.~(\ref{be4})-(\ref{be3}) outside 
the horizon by using the boundary conditions (\ref{f1a1})-(\ref{h2}) around $r=r_h$. In Fig.~\ref{fig1} we plot the numerically integrated solutions of $f, h,A_0, A_1,h-f$ for the coupling 
$\beta_3=7\times10^{-3}/(r_h M_{\rm pl})$. 
The two asymptotic solutions smoothly 
join each other without any singular behavior.
 While the metric $h$ approaches 1 as $r \to \infty$,  
the existence of the coupling $\beta_3$ in 
Eq.~(\ref{h2}) around the horizon leads to the value of $f$ 
slightly different from 1 in the large-distance limit.
We exploit the freedom of performing a time rescaling 
to shift $f$ to 1 by solving the equations of motion up to 
$r=10^{7}r_h$. After this rescaling, the difference 
between $f$ and $h$ is of the order of $10^{-2}$ 
around the horizon for the coupling $\beta_3$ chosen 
in Fig.~\ref{fig1}. For increasing $|\beta_3|$, 
this difference tends to be larger.
The deviation from GR is most significant in 
the regime of strong gravity and it starts to decrease for 
$r \gtrsim 10r_h$. This signature can be potentially 
probed by the future measurements of 
gravitational waves in the nonlinear regime of gravity.

From Fig.~\ref{fig1} we find that the longitudinal vector component behaves as $A_1 \propto (r-r_h)^{-1}$ around the horizon. As explained already, the apparent divergence of $A_1$ at $r=r_h$ does not spoil the regularity of the vector field. 
In the limit $r \to \infty$, we also numerically confirm that $A_1$ decreases 
in proportion to $r^{-2}$. The temporal vector component approaches 
the constant $P$ as $r \to \infty$. 

Due to the continuity of two asymptotic solutions outside 
the horizon, the model parameters $P,M,\tilde{b}_2$ in 
Eqs.~(\ref{G3f})-(\ref{G3A1}) should be related to 
the parameters $\mu, r_h, a_0$ in Eqs.~(\ref{f1a1})-(\ref{h2}), 
as $P=P(\mu, r_h, a_0)$, $M=M(\mu, r_h, a_0)$, and
$\tilde{b}_2=\tilde{b}_2(\mu, r_h, a_0)$. 
Since the constant $P$ cannot be fixed by other two parameters 
$M$ and $\tilde{b}_2$, it can be regarded as the primary Proca hair.

\subsection{$n \geq 2$}

Let us proceed to the cubic coupling (\ref{G3powermodel}) with 
the powers $n \geq 2$.
In this case, Eq.~(\ref{be5}) reduces to 
\be
\beta_3 
\left[ fh (rf'+4f)A_1^2+rA_0 (2fA_0'-f'A_0) \right]
\left( A_0^2 -fh A_1^2 \right)^{n-1}
=0\,. 
\label{be3nge2}
\ee
Hence, there are two branches characterized by 
(i) $fh (rf'+4f)A_1^2+rA_0 (2fA_0'-f'A_0)=0$ or 
(ii) $A_0^2 -fh A_1^2=0$, which we will 
discuss separately below.

\subsubsection{Branch {\rm (i)}}

For this branch, the longitudinal mode $A_1$ satisfies the 
relation same as Eq.~(\ref{A1be3}).
Taking the positive branch of $A_1$ for $r>r_h$ and
expanding $f,h,A_0$ around the horizon as Eq.~(\ref{fhA0}), 
we obtain the coefficients $f_1,h_1,a_1,f_2,h_2,a_2$ whose forms 
are the same as Eqs.~(\ref{f1a1})-(\ref{h2}) but with different 
values of ${\cal F}_2, {\cal H}_2, \alpha_2$. 
For $n=2$, they are given by 
\be
{\cal F}_2=\lambda  {\cal F}_2^{(n=1)}\,,\qquad 
{\cal H}_2 =\lambda  {\cal H}_2^{(n=1)}\,,\qquad 
\alpha_2=\lambda \alpha_2^{(n=1)}\,,
\ee
where ${\cal F}_2^{(n=1)}$, ${\cal H}_2^{(n=1)}$, and 
$\alpha_2^{(n=1)}$ correspond to the values of  
${\cal F}_2, {\cal H}_2, \alpha_2$ for $n=1$, and 
\be
\lambda=\frac{2a_0(\sqrt{2\mu}M_{\rm pl}+2 a_0)}
{(1-\mu)M_{\rm pl}^2}\,.
\ee
The effect of the coupling $\beta_3$ works in a similar way to 
that discussed for $n=1$. For larger $\beta_3$, the difference 
between the metrics $f$ and $h$ tends to be more significant
around the horizon.
The longitudinal mode around the horizon is given by 
Eq.~(\ref{A1ho}), so it behaves as $A_1 \propto (r-r_h)^{-1}$.

Far outside the horizon ($r \gg r_h$) we can also derive 
iterative solutions by using the expansion (\ref{fh}). 
Up to the order $1/r^4$, the solutions to $f,h,A_0$ for 
$n \geq 2$ are exactly the same as Eqs.~(\ref{G3f}), 
(\ref{G3h}), and (\ref{G3A0}), respectively.
At leading order, the longitudinal component 
decreases as $A_1=\tilde{b}_2/r^2$. 
Thus, the behavior of large-distance solutions 
is similar to that for $n=1$. The asymptotic solutions 
of $f,h,A_0,A_1$ for $r \gg r_h$ 
contain the three parameters $P,M,\tilde{b}_2$, which 
are related to other three parameters $\mu,r_h,a_0$ appearing 
in the solutions expanded around the horizon.
Since $P$ is not solely fixed by $M$ and $\tilde{b}_2$, 
the vector hair is of the primary type.

\subsubsection{Branch {\rm (ii)}}

The branch (ii) satisfies the relation  
\be
A_1=\epsilon \sqrt{\frac{A_0^2}{fh}}\,.
\label{A1spe}
\ee
Substituting Eq.~(\ref{A1spe}) 
into Eqs.~(\ref{be4}), (\ref{be1}), and (\ref{be2}), 
we obtain the differential equations same as 
Eqs.~(\ref{be3GR}), (\ref{be1GR}), and (\ref{be2GR}), 
respectively. Imposing the boundary conditions $f=h=1$ 
at $r \to \infty$, we obtain the RN metrics (\ref{RNso}) with 
the temporal vector component (\ref{RNso0}). 
Indeed, the relation (\ref{A1spe}) corresponds to the special 
case of Eq.~(\ref{A1A0re}) with $X_c=0$ and $f=h$. 
As we discussed in Sec.~\ref{G3subsec}, the RN solutions  
(\ref{RNso}) and (\ref{RNso0}) with 
the longitudinal mode (\ref{G3A1so}) exist for the theory 
given by the function (\ref{G3exactlag}) under the condition 
$G_{3,X}(X_c)=0$. Since we are now considering 
the coupling $G_3(X) \propto X^n$ with $n \geq 2$, 
the condition $G_{3,X}(X_c)=0$ is satisfied for 
$X_c=0$. Setting $X_c=0$ in Eq.~(\ref{G3A1so}), 
it follows that $A_1$ is equivalent to Eq.~(\ref{A1spe}) 
with $f,h,A_0$ given by Eqs.~(\ref{RNso}) and (\ref{RNso0}).

\section{Power-law quartic couplings $G_4(X)$}
\label{G4powersec}

Let us proceed to the model of quartic power-law 
interactions given by 
\be
G_4=\frac{M_{\rm pl}^2}{2}
+\beta_4 M_{\rm pl}^2\left( 
\frac{X}{M_{\rm pl}^2} \right)^n\,,
\label{G4power}
\ee
where $\beta_4$ is a dimensionless constant.
{}From Eq.~(\ref{be5}) the longitudinal mode obeys 
\be
\beta_{4}A_{1}
\left( A_0^{2} -fh A_{1}^{2}  \right) ^{n-2}
[A_1^2 fh \{ (1+h-2nh)f+(1-2n)rf'h \}+
A_0^2 \{ f(h-1)+(2n-1)rf'h \}-4r(n-1)A_0A_0'fh]=0\,.
\label{G4A1eq}
\ee

For $n=1$, Eq.~(\ref{G4A1eq}) reduces to 
\be
\beta_4 \left[ hrf'+(h-1)f \right]A_1=0\,,
\ee
so we have two branches satisfying  
(i) $hrf'+(h-1)f=0$ or (ii) $A_1=0$. 
As we showed in Sec.~\ref{exactsec}, there exists an exact 
BH solution for the branch (i) with $\beta_4=1/4$. 
In Ref.~\cite{Chagoya2} the solutions for general $\beta_4$ 
were discussed for the two branches (i) and (ii), 
so we will not repeat the analysis here.

In the following we will study the $n=2$ (vector Galileons) 
and $n \geq 3$ cases, separately.

\subsection{$n=2$}

In this case, Eq.~(\ref{G4A1eq}) yields
\be
\beta_4 A_1 \left[ A_1^2 fh \left\{ (1-3h)f-3rf'h \right\}
+A_0^2 \left\{ f(h-1)+3rf'h \right\}-4rA_0A_0' fh 
\right]=0\,.
\label{A1n=2}
\ee
Since there are two branches characterized by 
$A_1 \neq 0$ or $A_1=0$, we will 
discuss such two cases in turn.

\subsubsection{Branch with $A_1 \neq 0$}

This branch corresponds to the longitudinal mode satisfying 
\be
A_1=\epsilon \sqrt{\frac{A_0^2 \left\{ f(h-1)+3rf'h \right\}
-4rA_0A_0' fh}{fh[3rf'h-(1-3h)f]}}\,.
\label{G4A1}
\ee
We first derive the solutions around the horizon by expanding the 
functions $f,h,A_0$ in the forms (\ref{fhA0}) with the constraint (\ref{G4A1}). 
We take the positive branch of Eq.~(\ref{G4A1}) for $r>r_h$,  
assume that $f_1>0$ and $h_1>0$, and pick up terms 
linear in $\beta_4$. The resulting coefficients up to the 
order of $(r-r_h)^2$, which recover the 
RN metrics (\ref{fRN}) in the limit $\beta_4 \to 0$, read
\be
f_1=h_1=\frac{1-\mu}{r_h}\,,\qquad 
a_1= \frac{\sqrt{2\mu}M_{\rm pl}}{r_h}
+\alpha_1 \beta_4\,,
\label{f1a1d}
\ee
and 
\be
f_2 = \frac{2\mu-1}{r_h^2}
+{\cal F}_2 \beta_4\,,\qquad
h_2 = \frac{2\mu-1}{r_h^2}
+{\cal H}_2 \beta_4\,,\qquad
a_2 = -\frac{\sqrt{2\mu}M_{\rm pl}}{r_h^2} 
+\alpha_2 \beta_4\,,
\label{h2G4}
\ee
where 
\ba
\hspace{-1.1cm}
&&\alpha_1 =
\frac{3a_0^2\left[\sqrt{2\mu}(8\mu M_{\rm pl}^2+a_0^2)
+8\mu a_0M_{\rm pl}\right]}{(3\mu-2)^2M_{\rm pl}^3}
\,,\\
\hspace{-1.1cm}
&&{\cal F}_2 = 
\frac{4\mu a_0\left[2\sqrt{2\mu} M_{\rm pl} \left\{15a_0^2+4(3\mu-2) M_{\rm pl}^2\right\} 
+3a_0 \left\{3a_0^2+2(11\mu-2)M_{\rm pl}^2\right\}
\right]}{(3\mu-2)^2M_{\rm pl}^4r_h^2}
\,,\\
\hspace{-1.1cm}
&&{\cal H}_2 =
\frac{4\mu a_0\left[2\sqrt{2\mu} M_{\rm pl} \left\{3(14-9\mu)a_0^2-4(3\mu-2)(3\mu-4)M_{\rm pl}^2\right\}
+3a_0\left\{3(2-\mu)a_0^2-2(27\mu^2-40\mu+4)M_{\rm pl}^2\right\}
\right]}{(3\mu-2)^3M_{\rm pl}^4r_h^2}
\,,\\
\hspace{-1.1cm}
&&\alpha_2 =
-\frac{a_0\left[3\sqrt{2\mu}a_0\left\{96\mu^2M_{\rm pl}^2+(15\mu-2)a_0^2\right\}
+16 \mu M_{\rm pl} \left\{18\mu a_0^2+(9\mu^2-4) M_{\rm pl}^2\right\}
\right]}{(3\mu-2)^3M_{\rm pl}^3r_h^2}
\,.
\ea
Apart from the appearance of the $\beta_4$-dependent term in $a_1$, 
the structure of solutions around the horizon is similar to that of the 
power-law cubic models studied in Sec.~\ref{G3powersec}. 
The coupling $\beta_4$ works as corrections to the leading-order 
RN solutions characterized by the first terms on the 
r.h.s. of Eq.~(\ref{h2G4}). 
Taking the positive branch, 
the behavior of the longitudinal mode 
(\ref{G4A1}) is given by 
\be
A_1=\frac{a_0}
{f_1(r-r_h)}
+\frac{a_0[(f_2+h_2)(1-3f_1r_h)-2f_1^2]
+2a_1f_1(f_1r_h-1)}{2f_1^2 (3f_1r_h-1)}
+{\cal O}(r-r_h)\,.
\ee
The leading-order contribution to $A_1$ is similar to 
that in Eq.~(\ref{A1ho}) of the cubic-coupling case, so 
the regularity of solutions is ensured around the horizon. 

For the distance $r$ much lager than $r_h$, we perform 
the expansions of $f, h, A_0$ given by Eq.~(\ref{fh}).
In doing so, we take the $r$ derivative of Eq.~(\ref{G4A1}) and 
eliminate the terms $A_1$ and $A_1'$ from 
Eqs.~(\ref{be4}), (\ref{be1}), and (\ref{be2}). 
Picking up the leading-order terms of $\beta_4$, 
we obtain the iterative solutions 
\ba
\hspace{-0.2cm}
f &=& 1-\frac{2M}{r}+ \left[ 
\frac{Q^2}{2M_{\rm pl}^2}+\frac{3P^2Q^2 
(5P^2-8M_{\rm pl}^2) \beta_4}{4M_{\rm pl}^6} 
\right]\frac{1}{r^2}
+\frac{PQ^3 (3P^2-4M_{\rm pl}^2) \beta_4}
{M_{\rm pl}^6 r^3}+{\cal O} \left( \frac{1}{r^4} 
\right)\,,\label{fG4}\\
\hspace{-0.2cm}
h &=& 1-\frac{2M}{r}+ \left[ 
\frac{Q^2}{2M_{\rm pl}^2}+\frac{3P^2Q^2 
(11P^2-16M_{\rm pl}^2) \beta_4}{4M_{\rm pl}^6} 
\right]\frac{1}{r^2}
+\frac{PQ^2 (Q-3MP)(3P^2-4M_{\rm pl}^2) \beta_4}
{M_{\rm pl}^6 r^3}+{\cal O} \left( \frac{1}{r^4} 
\right), \label{hG4}\\
\hspace{-0.2cm}
A_0 &=& P+\frac{Q}{r}+\frac{PQ^2 
(3P^2-4M_{\rm pl}^2)\beta_4}{M_{\rm pl}^4r^2}
-\frac{Q^3 (3P^2-4M_{\rm pl}^2)^2\beta_4}
{12M_{\rm pl}^6r^3} +{\cal O} \left( \frac{1}{r^4} 
\right)\,,\label{A0G4}\\
\hspace{-0.2cm}
A_1 &=&
\frac{\sqrt{2P(MP+Q)}}{\sqrt{r}} 
\left[1+
\frac{2M_{\rm pl}^4\{8M_{\rm pl}^2(2MP+Q)^2-5P^2Q^2\}
+\beta_4 P^2Q^2\{64M_{\rm pl}^2(3P^2-2M_{\rm pl}^2)-57 P^4\}}
{32P(MP+Q)M_{\rm pl}^6r}\right]\notag\\
&&+{\cal O} \left( \frac{1}{r^{5/2}} \right)\,,
\label{A1G4}
\ea
where we have set $f_1=h_1=-2M$ and $\tilde{a}_1=Q$. 
We recover the RN solutions (\ref{RNso})-(\ref{RNso0}) 
by taking the limit $\beta_4 \to 0$ in 
Eqs.~(\ref{fG4})-(\ref{A0G4}).
The existence of the coupling $\beta_4$ leads to the difference 
between the two metric components $f$ and $h$ at the order of $1/r^2$. 
The leading-order longitudinal mode decreases as 
$A_1 \propto 1/\sqrt{r}$, whose property is 
different from that in the cubic power-law models 
(in which case $A_1 \propto r^{-2}$).

\begin{figure}
\begin{center}
\includegraphics[height=3.5in,width=3.5in]{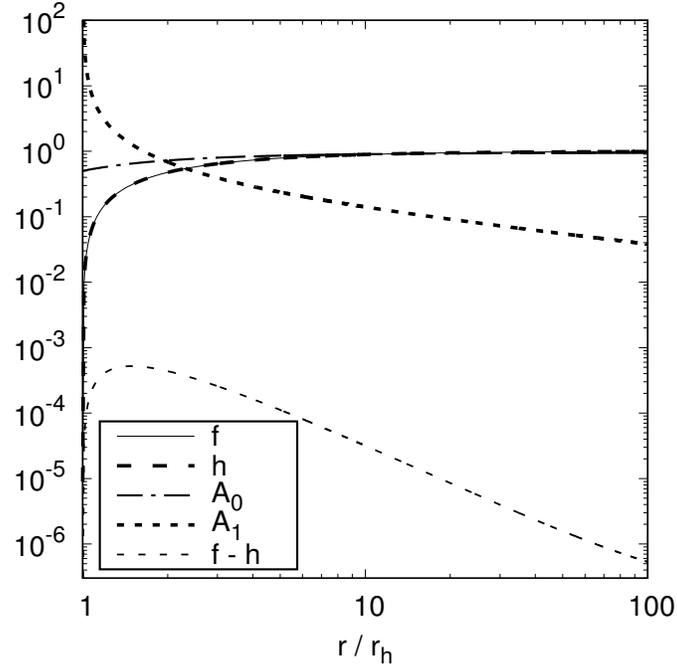}
\end{center}
\caption{\label{fig2}
Numerical solutions of $f, h, A_0, A_1, f-h$ outside the horizon 
for the quartic vector Galileon model 
$G_4(X)=\beta_4X^2/M_{\rm pl}^2$ 
with the coupling $\beta_4=10^{-2}$. 
We choose the boundary conditions (\ref{f1a1d})-(\ref{h2G4}) 
at $r=1.001r_h$ with $\mu=0.1$ and $a_0=0.5M_{\rm pl}$. 
The solutions around $r=r_h$ smoothly connect to those  
at spatial infinity.}
\end{figure}

In Fig.~\ref{fig2} we show the numerically integrated solutions 
to $f, h, A_0, A_1, f-h$ for $\beta_4=10^{-2}$ derived by 
choosing the boundary conditions (\ref{fhA0}) 
with Eqs.~(\ref{f1a1d})-(\ref{h2G4}) near the horizon.
The solutions smoothly connect to those in another 
asymptotic regime $r \gg r_h$. 
As in the numerical simulation of Fig.~\ref{fig1}, the metric $f$ 
is rescaled to 1 at a sufficiently large distance ($r=10^{7}r_h$).
For $\beta_4=10^{-2}$ the maximum difference 
between $f$ and $h$ is of the order of $10^{-3}$ 
around the horizon, but it decreases for larger $r$ according 
to Eqs.~(\ref{fG4})-(\ref{hG4}). 
Numerically, we also confirm that the longitudinal mode behaves as 
$A_1 \propto 1/(r-r_h)$ around $r \simeq r_h$ and 
$A_1 \propto 1/\sqrt{r}$ for $r \gg r_h$.
The parameters $ \mu, r_h, a_0$ around the horizon are related 
to $P, Q, M$ in Eqs.~(\ref{fG4})-(\ref{A1G4}). Since the constant 
$P$ can not be fixed by $Q$ and $M$, it is regarded as a primary hair.

\subsubsection{Branch with $A_1=0$}

{}From Eq.~(\ref{A1n=2}) there exists the other branch satisfying $A_1=0$. Substituting $A_1=0$ and $A_1'=0$ 
into Eqs.~(\ref{be4})-(\ref{be2}) and expanding the 
functions $f,h,A_0$ as Eq.~(\ref{fhA0}) around the 
horizon, the solutions recovering the RN metric (\ref{fRN}) 
in the limit $\beta_4 \to 0$ are given by 
\ba
f &=& (1-\mu) \Delta x-(1-2\mu)(\Delta x)^2 
+\left[1-3\mu-\frac{2\mu^2(6-7\mu)}{3(1-\mu)^2}\beta_4 
\right](\Delta x)^3+{\cal O}((\Delta x)^4)
\,,\label{fA1=0}\\
h &=& (1-\mu) \Delta x-(1-2\mu)(\Delta x)^2 
+\left[1-3\mu-\frac{10\mu^3}{3(1-\mu)^2}\beta_4 
\right](\Delta x)^3+{\cal O}((\Delta x)^4)\,,
\label{hA1=0}\\
A_0 &=& 
\sqrt{2\mu}M_{\rm pl} \left[ \Delta x-(\Delta x)^2
+\left\{ 1+\frac{2\mu^3}{3(1-\mu)^3} \beta_4 
\right\}(\Delta x)^3 \right]+{\cal O}((\Delta x)^4)\,,
\label{A0A1=0}
\ea
where 
\be
\Delta x = \frac{r}{r_h}-1\,,
\ee
and we have chosen the branch $a_1>0$. 
The temporal component $A_0$ exactly 
vanishes at $r=r_h$.
The effect of $\beta_4$ on $f,h,A_0$ arises 
at the order of $(\Delta x)^3$, which is higher than the 
order of the branch $A_1 \neq 0$.

For the distance $r \gg r_h$, the expansions of $f,h,A_0$ 
in the forms (\ref{fh}) give rise to the following iterative 
solutions (up to linear order in $\beta_4$):
\ba
f &=& 1-\frac{2M}{r} \left[ 1+\frac{2P^3 (MP+Q)\beta_4}
{MM_{\rm pl}^4} \right]+\frac{1}{r^2} 
\left[ \frac{Q^2}{2M_{\rm pl}^2} 
-\frac{4P^2\beta_4}{M_{\rm pl}^4} 
\left( M^2 P^2+2MPQ+\frac32 Q^2  
-\frac{5P^2 Q^2 }{16M_{\rm pl}^2}\right) \right]
+{\cal O} \left( \frac{1}{r^3} \right),\label{fA1=0d} 
\nonumber \\
& &\\
h &=& 
1-\frac{2M}{r}+\frac{1}{r^2} \left( \frac{Q^2}{2M_{\rm pl}^2}
+\frac{3P^4Q^2 \beta_4}{4M_{\rm pl}^6} \right)
+{\cal O} \left( \frac{1}{r^3} \right)\,, \label{hA1=0d}\\
A_0 &=&
P+\frac{Q}{r}-\frac{P^3 Q (2MP+Q)\beta_4}
{2M_{\rm pl}^4 r^2}+{\cal O} \left( \frac{1}{r^3} \right)\,,
\label{A0A1=0d}
\ea
where we have set $\tilde{a}_1=Q$ and $\tilde{h}_1=-2M$. 
The coupling $\beta_4$ works as corrections to the RN solutions (\ref{RNso})-(\ref{RNso0}), which induce 
the difference between $f$ and $h$ even at the order of $1/r$.
If we consider the case in which the conditions $|Q| \ll |MP|$ and $|\beta_4| \ll 1$ are satisfied,  
the horizon radius $r_h$ is close to $2M$. 
For $|P|$ of the order of $M_{\rm pl}$, Eqs.~(\ref{fA1=0d}) 
and (\ref{hA1=0d}) show that $|f-h|$ is of the order of 
$(r_h/r)\beta_4$ for $r \gg r_h$. This means that, even if 
the corrections induced by $\beta_4$ are small in 
the very vicinity of the horizon, they are not necessarily negligible for $r$ larger than the order of $r_h$ 
(say, $r =10r_h$). At spatial infinity the effects of $\beta_4$
on $f$ and $h$ are suppressed.

Besides the coupling $\beta_4$, we have two parameters $\mu$ and $r_h$ in Eqs.~(\ref{fA1=0})-(\ref{A0A1=0}), whereas 
there are three parameters $P,M,Q$ in 
Eqs.~(\ref{fA1=0d})-(\ref{A0A1=0d}). 
Numerically we confirmed that the two asymptotic solutions 
in the regimes $r \simeq r_h$ and $r \gg r_h$ smoothly 
join each other.
Hence $P,Q,M$ are related to $\mu, r_h$ according to 
$P=P(\mu,r_h)$, $Q=Q(\mu,r_h)$, and $M=M(\mu,r_h)$.
In this case $P$ depends on $Q$ and $M$, 
so the vector hair is of the secondary type. 

\subsection{$n \geq 3$}

For the theories with $n \geq 3$, there are the three 
branches of solutions:
\ba
{\rm (i)}~
& &A_1=\epsilon \sqrt{\frac{A_0^2 [f(h-1)+(2n-1)rf'h]
-4rA_0A_0'(n-1)fh}{fh[(2n-1)rf'h-(1+h-2nh)f]}}\,,\\
{\rm (ii)}~
& &A_1=0\,,\\
{\rm (iii)}~
& &A_1=\epsilon \sqrt{\frac{A_0^2}{fh}}\,.
\ea
The first two branches are similar to those discussed for 
$n=2$, but the branch (iii) arises only for $n \geq 3$.

For the branch (i), the solution around the horizon is given 
by Eq.~(\ref{fhA0}) with the coefficients in the forms 
(\ref{f1a1d}) and (\ref{h2G4}) but with different values of 
$\alpha_1, {\cal F}_2, {\cal H}_2, \alpha_2$. 
As in the case of $n=2$, the effect of the coupling $\beta_4$ 
induces the difference between $f$ and $h$ at the order 
of $(r-r_h)^2$. At spatial infinity
the modifications to the RN solutions arise at the 
order of $1/r^2$, whose property is similar to 
Eqs.~(\ref{fG4})-(\ref{A0G4}) of the $n=2$ case. 
If $n=3$, for example, the leading-order corrections to 
$f, h, A_0$ of the RN solutions (\ref{RNso})-(\ref{RNso0})  
are given, respectively, by 
\be
\delta f=\frac{5P^4Q^2(7P^2-12M_{\rm pl}^2)\beta_4}
{8M_{\rm pl}^8r^2}\,,\qquad
\delta h=\frac{5P^4Q^2(29P^2-48M_{\rm pl}^2)\beta_4}
{16 M_{\rm pl}^8r^2}\,,\qquad
\delta A_0=\frac{3P^3Q^2(5P^2-8M_{\rm pl}^2)\beta_4}
{4M_{\rm pl}^6 r^2}\,.
\ee
The longitudinal mode decreases as $A_1 \propto 1/\sqrt{r}$ for $r \gg r_h$. Our numerical simulations show that 
the solutions in two asymptotic regimes $r \approx r_h$ and 
$r \gg r_h$ smoothly join each other.

For the branch (ii), the effect of the coupling $\beta_4$ on 
the solution (\ref{fhA0}) expanded around the horizon 
appears at higher orders with increasing $n$, 
e.g., at the order of $(r-r_h)^4$ for $n=3$. 
At spatial infinity, the coupling $\beta_4$ leads 
to the difference between $f$ and $h$ at the order of $1/r$. 
If $n=3$, for example, the two 
metric components
are given by 
\be
f=1-\frac{2M}{r} \left[ 1+\frac{3P^5(MP+Q) \beta_4}
{2MM_{\rm pl}^6} \right]+{\cal O} 
\left( \frac{1}{r^2} \right)\,,\qquad 
h=1-\frac{2M}{r}+{\cal O} \left( \frac{1}{r^2} \right)\,,
\ee
so that $|f-h|$ can be of the order of $(r_h/r)\beta_4$ 
for $|Q| \ll |MP|$ and $|P|={\cal O}(M_{\rm pl})$.

For the branch (iii), it follows that Eqs.~(\ref{be4}), (\ref{be1}), and (\ref{be2}) reduce to Eqs.~(\ref{be3GR}), (\ref{be1GR}), and 
(\ref{be2GR}), respectively. 
Hence this branch corresponds to the RN solutions with 
$A_1=\epsilon A_0/f$.
Since $X=0$ in this case, we have $G_{4,X}=0$ and 
$G_{4,XX}=0$ for $n \geq 3$. This is the reason why 
all the $X$-dependent terms arising from $G_4(X)$ 
vanish from the equations of motion. 
Note that the exact solution discussed in Sec.~\ref{exactsec} 
is different from the above non-exact solution, because the former 
satisfies the condition $G_{4,X}=1/4$.

\section{Power-law quintic couplings $G_5(X)$}
\label{G5powersec}

In this section, we consider power-law quintic couplings 
given by 
\be
G_5=\beta_5 \left( \frac{X}{M_{\rm pl}^2} \right)^n\,,
\label{G5coupling}
\ee
with $G_4=M_{\rm pl}^2/2$,  
where $\beta_5$ is a dimensionless constant.
{}From Eq.~(\ref{be5}) it follows that 
\ba
& &
\beta_5 \left( A_0^2 -fh A_1^2 \right)^{n-2} 
[ (2nh+h-1)f^2h^2 f' A_1^4 
-2fh \{ (nh+h-1)A_0f'+(1+h-2nh)A_0' f \}A_0A_1^2
\nonumber \\
& &
+(h-1)(A_0f'-2fA_0')A_0^3 ]=0\,.
\label{A1eqbe5}
\ea

If $n=1$, then this reduces to 
\be
A_1=\epsilon \sqrt{\frac{A_0(h-1)(A_0f'-2A_0'f)}
{ff'h(3h-1)}}\,,
\label{A1G5n=1}
\ee
so there are two physically equivalent branches 
corresponding to $\epsilon=\pm 1$.
 
For $n \geq 2$, it follows that 
\be
A_1=\epsilon \sqrt{\xi_1 \left[ 1 \pm \sqrt{1-\frac{\xi_2}{\xi_1^2}} 
\right]}\,,\label{A1ngeq2}
\ee
where 
\be
\xi_1 = \frac{A_0[\{ (n+1)h-1 \}A_0f'
+\{ 1+(1-2n)h \}A_0' f]}
{ff'h [(2n+1)h-1]}\,,\qquad
\xi_2 = \frac{(h-1)A_0^3(A_0f'-2A_0'f)}
{f^2h^2f'  [(2n+1)h-1]}\,.\label{xi2}
\ee
We require the two conditions $\xi_1^2 \geq \xi_2$ and 
$\xi_1[1 \pm \sqrt{1-\xi_2/\xi_1^2}] \geq 0$
for the existence of the solutions (\ref{A1ngeq2}).

For $n \geq 3$, Eq.~(\ref{A1eqbe5}) admits the following solutions
\be
A_1=\epsilon \sqrt{\frac{A_0^2}{fh}}\,, 
\label{A1G5}
\ee
besides the solutions satisfying Eq.~(\ref{A1ngeq2}). 
As in the cases of cubic and quartic power-law couplings, 
the branches (\ref{A1G5}) correspond to the RN solutions 
(\ref{RNso}) and (\ref{RNso0}). 

In what follows, we will focus on the branches (\ref{A1G5n=1}) 
and (\ref{A1ngeq2}) for $n=1$ and $n \geq 2$, respectively. 
At spatial infinity, we expand $f,h,A_0$ in the forms (\ref{fh}) and also assume the asymptotic behavior
$A_1=\sum_{i=1}^{\infty} \tilde{b}_i/r^i$.
Then, we obtain the iterative solutions 
\ba
f&=& 1-\frac{2M}{r}
+\frac{M^2 P^2}{2M_{\rm pl}^2 r^2}
+\epsilon \frac{
2^{1-n}P^{1+2n}n\,M^2 
\beta_5\sqrt{2\mu_P}}{3M_{\rm pl}^{3+2n}r^3}
+{\cal O}\left( 
\frac{1}{r^4} \right)\,,\label{G5fd} \\
h&=& 1-\frac{2M}{r}
+\frac{M^2 P^2}{2M_{\rm pl}^2 r^2}
+\epsilon \frac{
2^{1-n}P^{1+2n}n\,M^2 
\beta_5\sqrt{2\mu_P}}
{M_{\rm pl}^{3+2n} r^3}
+{\cal O}\left( 
\frac{1}{r^4} \right)\,,\label{G5hd} \\
A_0&=& 
P-\frac{PM}{r}
+\epsilon\frac{
2^{1-n}P^{2n}n\,M^2 
\beta_5\sqrt{2\mu_P}}{3M_{\rm pl}^{1+2n}r^3}
+{\cal O} \left( 
\frac{1}{r^4} \right)\,,\label{G5A0d} \\
A_1&=&
\epsilon \frac{MP \sqrt{2\mu_P}}
{2M_{\rm pl}r}+\frac{MP}{M_{\rm pl}^3r^2} 
\biggl[ \frac{\epsilon}{4} \left\{ 
\frac{2n-1}{4} P^2
-(n-7)M_{\rm pl}^2 \right\}M
\sqrt{2\mu_P} 
+\frac{n P^{2n-1}\mu_P \beta_5}
{2^n M_{\rm pl}^{2n-1}} \biggr]
+{\cal O} \left( 
\frac{1}{r^3} \right)\,,
\label{G5A1d}
\ea
where $\mu_P=P^2-2M_{\rm pl}^2$, and 
we set $\tilde{a}_0=P>0$ and 
$\tilde{f}_1=\tilde{h}_1=-2M$. 
The existence of the above solutions requires 
the condition $\mu_P \geq 0$, i.e., 
$P^2 \geq 2M_{\rm pl}^2$. 
On using the large-distance solutions (\ref{G5fd})-(\ref{G5A0d}), 
we find that 
the quantities $\xi_1$ and $\xi_2$ defined by Eq.~(\ref{xi2}) 
behave as
\be
\xi_1 \simeq \frac{P^2}{2n}\,,\qquad 
\xi_2 \simeq \frac{M^2P^4 (P^2-2M_{\rm pl}^2)}
{2nM_{\rm pl}^2r^2}\,.
\ee
For the branch with the positive sign inside the square root 
of  Eq.~(\ref{A1ngeq2}),  we have that $A_1^2 \simeq 2\xi_1 \simeq P^2/n
={\rm constant}$ at spatial infinity.
The branch with the negative sign inside the square root of Eq.~(\ref{A1ngeq2}) gives rise to the solution 
$A_1^2 \simeq \xi_2/(2\xi_1) \simeq 
M^2P^2\mu_P/(2M_{\rm pl}^2 r^2)$, 
so this corresponds to the leading-order solution of Eq.~(\ref{G5A1d}). 

Outside the horizon the metric $h$ is in the range $0<h<1$. 
For $n=1$, Eq.~(\ref{A1G5n=1}) shows that $A_1^2$ 
exhibits the divergence at 
\be
h=\frac13\,.
\ee
For the $\epsilon=-1$ branch of Eq.~(\ref{A1G5n=1})
we can derive the solutions in the form (\ref{fhA0}) 
expanded around the horizon. On using such analytic 
boundary conditions and solving the equations 
of motion numerically, we find that $A_1$ indeed diverges 
as $h$ approaches $1/3$. 
Hence the solutions in the strong-gravity regime 
($h<1/3$) are disconnected to the large-distance solutions 
(\ref{G5fd})-(\ref{G5A1d}).

{}From Eqs.~(\ref{A1ngeq2})-(\ref{xi2}) we find that the similar 
divergence of $A_1$ occurs at 
\be
h=\frac{1}{2n+1}\,,
\label{hngeq2}
\ee
for $n \geq 2$. Since the metric (\ref{hngeq2}) is in the range $0<h \leq 1/5$, 
the solutions outside the horizon cannot avoid passing through 
this divergent point. Our numerical simulations show that there are 
no regular exterior BH solutions that smoothly connect to 
Eqs.~(\ref{G5fd})-(\ref{G5A1d}).

\section{Power-law sixth order couplings $G_6(X)$}
\label{G6powersec}

We proceed to the case of power-law sixth-order 
interactions given by 
\be
G_6=\frac{\beta_6}{M_{\rm pl}^2} 
\left( \frac{X}{M_{\rm pl}^2} \right)^n\,,
\label{G6coupling}
\ee
with $G_4=M_{\rm pl}^2/2$, where $\beta_6$ is a dimensionless constant.
The $U(1)$-invariant gravitational coupling advocated
by Horndeski \cite{Horndeski76} corresponds to $n=0$, 
so we will also include such a case in the analysis.
{}From Eq.~(\ref{be5}) the longitudinal mode obeys 
\be
\beta_6 A_0'^2{\cal G}A_1=0\,, 
\label{be6A1d}
\ee
where 
\be
{\cal G} \equiv 
 \left( A_0^2 -fh A_1^2 \right)^{n-2}
\left[ A_1^2 fh \{ (2n+1)h-1 \}-A_0^2 (3h-1) 
\right]\,. \label{calF}
\ee
Equation (\ref{be6A1d}) admits the solution $A_0'=0$, 
but this corresponds to the stealth Schwarzschild 
solution (\ref{fhG6}).
This is analogous to the discussion given in 
Sec.~\ref{G6A0v}, but the difference is that $A_1$ 
is arbitrary in the present case 
(since we are not imposing the condition that 
$X$ is constant). There exist other two branches 
satisfying ${\cal G}=0$ or $A_1=0$. 
Let us first discuss the possibility for the realization 
of the branch ${\cal G}=0$.

For $n=0$ the branch ${\cal G}=0$ is realized for 
$A_1^2/A_0^2=(3h-1)/[fh(h-1)]$, but the real solutions 
to $A_1$ do not exist for $1/3<h<1$.
When $n=1$ the quantity (\ref{calF}) simply reduces to 
${\cal G}=1-3h$, so there is no consistent branch satisfying
${\cal G}=0$ in the whole region outside the horizon. 
For $n=2$ we have
${\cal G}=A_1^2 fh(5h-1)-A_0^2(3h-1)$, so the 
real solutions to ${\cal G}=0$ are not present for
$1/5<h<1/3$. 
For $n \geq 3$ there are two solutions to ${\cal G}=0$, i.e., 
\be
{\rm (i)}~\,\frac{A_1^2}{A_0^2}= 
\frac{3h-1}{fh[(2n+1)h-1]}\,, \qquad \quad
{\rm (ii)}~\,A_1=\epsilon \sqrt{\frac{A_0^2}{fh}}\,.
\ee
The branch (i) does not exist in the region $1/(2n+1)<h<1/3$ 
outside the horizon. 
For the branch (ii) the solutions are described by the RN solutions (\ref{RNso})-(\ref{RNso0}).
Therefore, apart from the trivial branch (ii) present for $n \geq 3$, 
there are no consistent solutions satisfying ${\cal G}=0$.

Since the remaining possibility is the branch 
\be
A_1=0\,,
\ee
we will focus on this case in the following discussion. 
Substituting $A_1=0$ and $A_1'=0$ into Eqs.~(\ref{be4}), 
(\ref{be1}), (\ref{be2}) and expanding $f,h,A_0$ 
in the forms (\ref{fh}) at large distances ($r \gg r_h$), 
the iterative solutions are given by 
\ba
f&=& 1-\frac{2M}{r}+\frac{Q^2}{2M_{\rm pl}^2 r^2}
-\frac{\beta_6 P^{2n}Q^2}
{2^{1+n}M_{\rm pl}^{4+2n}r^4}
-\frac{2^{-n}\beta_6 P^{2n-1}Q^2 
\left[M P (6n-5)+8Qn \right]}
{10M_{\rm pl}^{4+2n}r^5} +{\cal O}\left( 
\frac{1}{r^6} \right)\,,\label{f0G6}\\
h&=& 1-\frac{2M}{r}+\frac{Q^2}{2M_{\rm pl}^2 r^2}
+\frac{\beta_6 M P^{2n} Q^2(2n-1)}{2^{1+n}M_{\rm pl}^{4+2n}r^5}
+{\cal O}\left( 
\frac{1}{r^6} \right)\,,\\
A_0&=&
P+\frac{Q}{r}-\frac{2^{-n}\beta_6 M P^{2n}Q}
{M_{\rm pl}^{2+2n}r^4}
-\frac{2^{-n}\beta_6 P^{2n-1}Q
(32M^2M_{\rm pl}^2P n+28MM_{\rm pl}^2 Q\,n
-3P Q^2)}{20M_{\rm pl}^{4+2n}r^5}+{\cal O}\left( 
\frac{1}{r^6} \right)\,.
\label{A0P}
\ea
For $n=0$ these results match with those derived by 
Horndeski in Ref.~\cite{HorndeskiBH}.
The coupling $\beta_6$ works as corrections to the 
leading-order RN solutions. The difference between $f$ 
and $h$ arises at the order of $1/r^4$.

We expand the solutions around the horizon as Eq.~(\ref{fhA0}) and pick up the terms linear in $\beta_6$.
The resulting solutions, which recover the RN metrics in the 
limit $\beta_6 \to 0$, are given by 
\ba
f&=& (1-\mu) \Delta x+\left[ 2\mu-1+
\mu (1-3\mu)\frac{\beta_6}{r_h^2 M_{\rm pl}^2} \right]
(\Delta x)^2+{\cal O} ((\Delta x)^3)\,,\\
h&=& (1-\mu) \Delta x+\left[ 2\mu-1+
\mu (\mu-3)\frac{\beta_6}{r_h^2 M_{\rm pl}^2} \right]
(\Delta x)^2+{\cal O} ((\Delta x)^3)\,,\\
A_0&=& a_0+\sqrt{2\mu}M_{\rm pl} \left( 
1-\frac{\beta_6}{r_h^2 M_{\rm pl}^2} \right) \Delta x
-\sqrt{2\mu}M_{\rm pl}  \left( 
1-\frac{4\beta_6}{r_h^2 M_{\rm pl}^2} \right)(\Delta x)^2
+{\cal O} ((\Delta x)^3)\,,
\ea
for $n=0$, and 
\ba
f&=& (1-\mu) \Delta x+\left[ 2\mu-1+
\frac{\mu^2}{1-\mu}
\frac{\beta_6}{r_h^2 M_{\rm pl}^2} \right]
(\Delta x)^2+{\cal O} ((\Delta x)^3)\,,\label{fn=1}\\
h&=& (1-\mu) \Delta x+\left[ 2\mu-1
-\frac{3\mu^2}{1-\mu}
\frac{\beta_6}{r_h^2 M_{\rm pl}^2} \right]
(\Delta x)^2+{\cal O} ((\Delta x)^3)\,,\\
A_0&=& \sqrt{2\mu}M_{\rm pl} \Delta x
-\sqrt{2\mu}M_{\rm pl}  \left[
1-\frac{\mu^2}{(1-\mu)^2}\frac{\beta_6}{r_h^2 M_{\rm pl}^2} \right]
(\Delta x)^2+{\cal O} ((\Delta x)^3)\,,
\label{A0n=1}
\ea
for $n=1$. 
If $n \geq 2$, the effect of the coupling $\beta_6$ arises at the 
order of $(\Delta x)^{n+1}$ in $f, h, A_0$.

\begin{figure}
\begin{center}
\includegraphics[height=3.5in,width=3.5in]{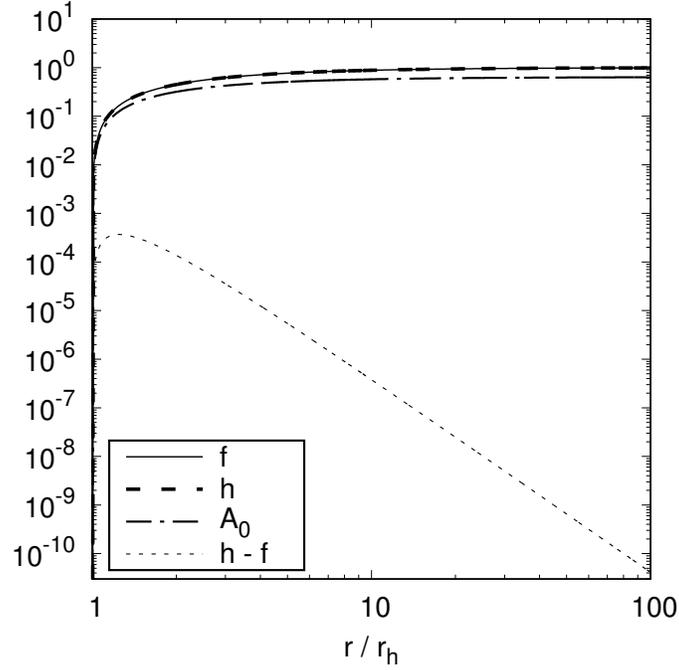}
\end{center}
\caption{\label{fig3}
Numerical solutions of $f, h,A_0,h-f$ outside the horizon 
for the sixth-order interaction
$G_6(X)=\beta_6X/M_{\rm pl}^4$ with the coupling 
$\beta_6=0.1r_h^2 M_{\rm pl}^2$. This corresponds 
to the branch $A_1=0$. We choose the boundary conditions 
(\ref{fn=1})-(\ref{A0n=1}) at $r=1.001r_h$ with $\mu=0.2$. The solutions are regular throughout the horizon exterior.}
\end{figure}

For $n=0$ there exists the $U(1)$ gauge symmetry, so 
the constant $P$ in Eq.~(\ref{A0P}) has no physical 
meaning with the value of $a_0$ unconstrained. 
In this case, we have two physical hairs $M$ and $Q$ 
related to the parameters $\mu$ and $r_h$ around the horizon.

For $n \geq 1$ we have that $a_0=0$, so the parameters 
$M,Q,P$ are related to the two parameters $\mu, r_h$ 
appearing for the solutions around the horizon.
Then, the Proca hair is of the secondary type.
This situation is analogous to what happens for the quartic 
power-law interactions with the branch $A_1=0$.

In Fig.~\ref{fig3} we plot the numerically integrated solutions 
to $f, h,A_0,h-f$ for $n=1$ with the branch $A_1=0$.
On choosing the boundary conditions (\ref{fn=1})-(\ref{A0n=1}) 
around $r=r_h$, the solutions smoothly connect to those 
in the regime $r \gg r_h$, i.e., Eqs.~(\ref{f0G6})-(\ref{A0P}) 
with $n=1$. Since $a_0=0$ in this case, the temporal 
component $A_0$ vanishes on the horizon. 
By normalizing the metric $f$ to be 1 at $r \to \infty$, 
the coupling $\beta_6$ induces the difference between 
$f$ and $h$ around the horizon. 
Compared to the cases of cubic and quartic couplings plotted 
in Figs.~\ref{fig1} and \ref{fig2}, $|f-h|$ decreases faster for 
increasing $r$. Thus, the future precise measurements for the 
deviation from GR in the strong-gravity regime may allow us to 
distinguish between hairy solutions with different couplings.
For $n=0$ we have also confirmed that the numerical solutions are 
regular outside the horizon with the difference between $f$ and $h$.
For $n \geq 2$, the effect of the coupling $\beta_6$ 
arises at higher order in metrics around the horizon.

\section{Power-law intrinsic vector mode couplings 
$g_4(X)$ and $g_5(X)$}
\label{g45powersec}

Let us finally study the models of power-law 
intrinsic vector-mode couplings given by 
\be
g_4(X)=\gamma_4 \left( \frac{X}{M_{\rm pl}^2} 
\right)^n\,,\qquad 
g_5(X)=\frac{\gamma_5}{M_{\rm pl}^2} 
\left( \frac{X}{M_{\rm pl}^2} \right)^n\,,
\label{ga45coupling}
\ee
where $g_4(X)$ is given in Eq.~\eqref{G2g4}, 
with $G_4=M_{\rm pl}^2/2$, where 
$\gamma_4$ and $\gamma_5$ are dimensionless constants.

\subsection{$\gamma_4 \neq 0$ and $\gamma_5=0$}

In this case, there is the relation (\ref{A02A1}) with 
$g_{4,X}=n\gamma_4 (A_0^2-fh A_1^2)^{n-1}
/[(2f)^{n-1}M_{\rm pl}^{2n}]$.
The branch satisfying $A_0'^2=0$ corresponds to 
the stealth Schwarzschild BH solution (\ref{fhG6}). 
For $n \geq 2$ there exists the branch $A_1=\epsilon \sqrt{A_0^2/(fh)}$, 
in which case the solutions are described by the RN
solutions (\ref{RNso})-(\ref{RNso0}).

In what follows we will focus on 
the last branch of 
Eq.~(\ref{A02A1}) with the vanishing longitudinal 
mode ($A_1=0$). 
Expanding the functions $f,h,A_0$ in the forms (\ref{fh}), 
the resulting large-distance solutions 
are given by 
\ba
f &=& 1-\frac{2M}{r}+\frac{1}{r^2} 
\left[ \frac{Q^2}{2M_{\rm pl}^2} 
-\frac{(n+1)P^{2n}Q^2 \gamma_4}
{2^n M_{\rm pl}^{2(n+1)}} \right]+{\cal O}\left( 
\frac{1}{r^3} \right)\,,\label{g4f} \\
h &=& 1-\frac{2M}{r}+\frac{1}{r^2} 
\left[ \frac{Q^2}{2M_{\rm pl}^2} 
-\frac{(2n+1)P^{2n}Q^2 \gamma_4}
{2^n M_{\rm pl}^{2(n+1)}}
 \right]+{\cal O}\left( 
\frac{1}{r^3} \right)\,,\\
A_0 &=& P+\frac{Q}{r} 
+\frac{n P^{2n-1}
Q(2MP+Q)\gamma_4}
{2^{n}(M_{\rm pl}^{2n}-2^{1-n} 
P^{2n} \gamma_4) r^2}+{\cal O}\left( 
\frac{1}{r^3} \right)\,.\label{A0g4}
\ea
For $n \geq 1$, the coupling $\gamma_4$ induces the difference 
between the two metrics $f$ and $h$ at the order of $1/r^2$.

For $n=1$, the solutions expanded around $r=r_h$, which recover
the RN metrics in the limit $\gamma_4 \to 0$, are given by 
\ba
f&=& (1-\mu) \Delta x+\left( 2\mu-1
-\frac{\mu^2}{1-\mu}\gamma_4 \right)
(\Delta x)^2+{\cal O} ((\Delta x)^3)\,,\\
h&=& (1-\mu) \Delta x+\left( 2\mu-1+
\frac{3\mu^2}{1-\mu}\gamma_4 \right)
(\Delta x)^2+{\cal O} ((\Delta x)^3)\,,\\
A_0&=& \sqrt{2\mu}M_{\rm pl} \Delta x
-\sqrt{2\mu}M_{\rm pl}  \left[ 
1+\frac{\mu^2}{(1-\mu)^2}\gamma_4 \right]
(\Delta x)^2
+{\cal O} ((\Delta x)^3)\,.
\ea
For $n \geq 2$, the effect of the coupling $\gamma_4$ 
appears at the order of $(\Delta x)^{n+1}$
in the expansions of $f,h,A_0$. 
Since $a_0=0$ for $n \geq 1$, there are two 
parameters $\mu$ and $r_h$ around the horizon. 
Since these two parameters are related to $P, Q, M$ in 
Eqs.~(\ref{g4f})-(\ref{A0g4}), the Proca hair $P$ 
is of the secondary type.

Numerically, we have confirmed that the two asymptotic solutions given above smoothly connect to each other. 
The coupling $\gamma_4$ gives rise to the difference between 
$f$ and $h$ in the strong-gravity regime, whose effect tends to 
be smaller for larger $n$. 

\subsection{$\gamma_5 \neq 0$ and $\gamma_4=0$}

For the quintic intrinsic vector-mode interaction, 
Eq.~(\ref{be5}) reduces to 
\be
\gamma_5 A_0'^2 \left( A_0^2-fh A_1^2 \right)^{n-1} 
\left[ A_0^2-(1+2n)fh A_1^2 \right]=0\,.
\ee
Again, the branch $A_0'^2=0$ corresponds to 
the stealth Schwarzschild solution (\ref{fhG6}). 
For $n \geq 2$ there exists the branch 
$A_1=\epsilon \sqrt{A_0^2/(fh)}$, which leads to 
the RN solutions (\ref{RNso})-(\ref{RNso0}).

\begin{figure}
\begin{center}
\includegraphics[height=3.5in,width=3.5in]{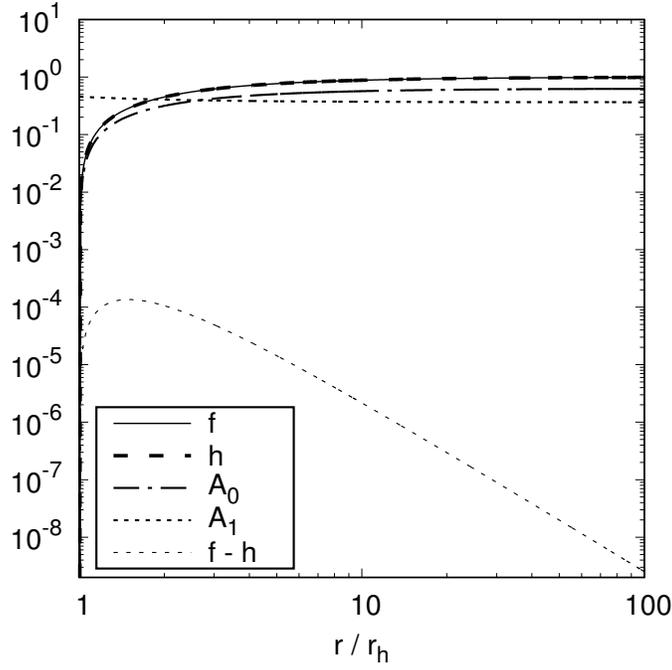}
\end{center}
\caption{\label{fig4}
Numerical solutions of $f, h,A_0,A_1,f-h$ outside the 
horizon for the quintic intrinsic vector-mode interaction
$g_5(X)=\gamma_5X/M_{\rm pl}^4$ with 
$\gamma_5=0.1 r_hM_{\rm pl}$. 
This corresponds to the branch (\ref{A1g5}) with 
$\epsilon=+1$. The boundary conditions around $r=r_h$
are chosen to be Eqs.~(\ref{fg6})-(\ref{A0g6}) with 
$\mu=0.2$ at $r=1.001r_h$. }
\end{figure}

For $n \geq 1$ we have the last branch satisfying 
\be
A_1=\epsilon \sqrt{\frac{A_0^2}{(1+2n)fh}}\,,
\label{A1g5}
\ee
where we will take the $\epsilon=+1$ branch in the following.
Differentiating Eq.~(\ref{A1g5}) with respect to $r$ and 
substituting $A_1$ and $A_1'$ into Eqs.~(\ref{be4}), (\ref{be1}), 
and (\ref{be2}), 
the iterative solutions at spatial infinity are given by 
\ba
f &=& 1-\frac{2M}{r}+\frac{Q^2}{2M_{\rm pl}^2r^2} 
-\frac{2n^{n+1}P^{2n+1}Q^2\gamma_5 }
{3(2n+1)^{n+1/2}M_{\rm pl}^{2(n+2)}\,r^3}
+{\cal O}\left( \frac{1}{r^4} \right)\,,\label{g5f} \\
h &=& 1-\frac{2M}{r}+\frac{Q^2}{2M_{\rm pl}^2r^2} 
-\frac{2n^{n+1}P^{2n+1}Q^2\gamma_5 }
{(2n+1)^{n+1/2}M_{\rm pl}^{2(n+2)}\,r^3}+{\cal O}\left( 
\frac{1}{r^4} \right)\,,\\
A_0 &=& P+\frac{Q}{r} 
+\frac{2n^n P^{2n+1}Q\gamma_5}
{(2n+1)^{n+1/2}M_{\rm pl}^{2(n+1)}r^2}
+\frac{n^n P^{2n}Q (8MPn+6Qn+3Q)\gamma_5}
{3M_{\rm pl}^{2(n+1)}(2n+1)^{n+1/2}\,r^3}
+{\cal O}\left( 
\frac{1}{r^4} \right)\,.
\label{A0g5}
\ea
The longitudinal mode behaves as
\be
A_1=\frac{P}{\sqrt{2n+1}}+\frac{2MP+Q}{\sqrt{2n+1}\,r}
+{\cal O}\left( \frac{1}{r^2} \right)\,,
\ee
which approaches the constant $P/\sqrt{2n+1}$ as $r \to \infty$.
The coupling $\gamma_5$ induces the difference between 
$f$ and $h$ at the order of $1/r^3$.
 
Expanding the solutions around $r=r_h$ as Eq.~(\ref{fhA0})
for $n \geq 1$, it follows that $a_0=0$. 
The effect of the coupling $\gamma_5$ arises at the order of 
$(r-r_h)^{n+2}$ in the expansions of $f,h,A_0$ as corrections 
to the leading-order RN solutions.
When $n=1$, for example, the resulting solutions are given by 
\ba
f&=& (1-\mu) \Delta x+\left( 2\mu-1 \right)(\Delta x)^2
+\left[ 1-3\mu+\frac{4\sqrt{6}\mu^{5/2}}{27(1-\mu)}
\frac{\gamma_5}{r_hM_{\rm pl}} \right] (\Delta x)^3
+{\cal O} ((\Delta x)^4)\,,\label{fg6} \\
h&=&  (1-\mu) \Delta x+\left( 2\mu-1 \right)(\Delta x)^2
+\left[ 1-3\mu+\frac{28\sqrt{6}\mu^{5/2}}{27(1-\mu)}
\frac{\gamma_5}{r_hM_{\rm pl}} \right] (\Delta x)^3
+{\cal O} ((\Delta x)^4)\,,\\
A_0&=& \sqrt{2\mu}M_{\rm pl} \Delta x
-\sqrt{2\mu}M_{\rm pl}  
(\Delta x)^2+\sqrt{2\mu}M_{\rm pl} \left[ 
1+\frac{2\sqrt{6}\mu^{3/2}(1-3\mu)}{27(1-\mu)^2}
\frac{\gamma_5}{r_hM_{\rm pl}} \right] (\Delta x)^3
+{\cal O} ((\Delta x)^4)\,.\label{A0g6}
\ea
There are two parameters $\mu$ and $r_h$ in the expansions 
(\ref{fg6})-(\ref{A0g6}), which are related to the three parameters $P,Q,M$ 
in Eqs.~(\ref{g5f})-(\ref{A0g5}). 
Hence the Proca hair $P$ is of the secondary type.

In Fig.~\ref{fig4} we plot the numerically integrated solutions 
for $n=1$ and  $\gamma_5=0.1 r_hM_{\rm pl}$ derived by using 
the boundary conditions (\ref{fg6})-(\ref{A0g6}) around $r=r_h$.
The solutions in two asymptotic regimes smoothly join each other. 
Substituting Eqs.~(\ref{fg6})-(\ref{A0g6}) into Eq.~(\ref{A1g5}), 
it follows that $A_1$ approaches the constant  
$M_{\rm pl}\sqrt{2\mu/[(1+2n)(1-\mu)^2]}$ as $r \to r_h$.
Indeed, the numerical simulation of Fig.~\ref{fig4} shows 
that $A_1$ starts to decrease from this finite value with increasing $r$ 
and it approaches another constant $P/\sqrt{2n+1}$ in the 
limit $r \to \infty$. {}From Eqs.~(\ref{fg6})-(\ref{A0g6}) the effect of 
the coupling $\gamma_5$ on $f,h, A_0$ does not arise up to 
the order of $(\Delta x)^3$,
but the difference between $f$ and $h$ still remains in the strong-gravity regime.

\section{Conclusions}
\label{concludesec}

In this paper, we have studied the static and spherically symmetric BH solutions in second-order generalized Proca 
theories with nonlinear derivative vector-field interactions.
In Sec.~\ref{modelsec} we derived the full background 
equations of motion for the action (\ref{action}) and 
revisited the non-existence of hairy BH solutions 
for a massive Proca field given by the Lagrangian 
$G_2=m^2X$.
More generally, we found that the bare 
$X$-dependent coupling $g_2(X)$ in $G_2$ without being multiplied by derivative terms like $F$ is generally the 
obstacle for the existence of hairy BH solutions. 
On the other hand, other derivative couplings like those 
appearing in the Lagrangians (\ref{L3})-(\ref{L6}) can 
give rise to a variety of hairy BH solutions.

In Sec.~\ref{exactsec} we reviewed the exact stealth BH solution which is known to exist for the specific coupling 
$X/4$ in $G_4$, and also obtained the extremal RN solution 
with the vanishing longitudinal mode (\ref{extremal})
present for the model (\ref{G4A10}). 
Imposing the two conditions of 
the two identical metric components (\ref{exactcon1}) 
and the constant norm of the vector field (\ref{exactcon2}), 
we also constructed a family of exact 
BH solutions for other interactions $G_3, G_5$ 
and intrinsic vector-mode couplings 
$G_6, G_2=-2g_4(X)F, g_5$ in Sec.~\ref{exactsec2}. 
The models allowing for their existence are given, 
respectively, by Eqs.~(\ref{G3exactlag}), (\ref{G5explicit}), 
(\ref{G6explicit}), (\ref{g4exact}), and (\ref{g5exact}).
The corresponding metrics  are described by either RN, extremal RN, or Schwarzschild types.

In Sec.~\ref{G3powersec} we explored the existence of
non-exact BH solutions for the power-law cubic interaction 
(\ref{G3powermodel}) with the non-vanishing longitudinal mode (\ref{A1be3}). 
Expanding $f,h,A_0$ around the 
horizon $r \simeq r_h$ as Eq.~(\ref{fhA0}) for $n \geq 1$, 
it follows that the coupling $\beta_3$ arises as corrections 
to the RN solutions. On using Eqs.~(\ref{f1a1})-(\ref{al2}) 
as the boundary conditions around $r=r_h$ and 
numerically solving the equations of motion outside the 
horizon, we showed that the solutions smoothly connect to 
those at spatial infinity given by Eqs.~(\ref{G3f})-(\ref{G3A1}). 
The constant $P$ appearing in Eq.~(\ref{G3A0}) cannot be fixed by other two parameters $M$ and $\tilde{b}_2$, so 
it corresponds to a primary hair.
As seen in Fig.~\ref{fig1}, the difference between two metrics $f$ and $h$ is most significant in the regime of strong gravity ($r \lesssim 10r_h$).

In Sec.~\ref{G4powersec} we showed that the power-law quartic coupling (\ref{G4power}) with $n \geq 2$ 
gives rise to two branches characterized by 
(i) $A_1 \neq 0$ with a primary Proca hair, and 
(ii) $A_1=0$ with a secondary Proca hair. 
In both cases, the solutions are regular throughout the 
horizon exterior with the difference between $f$ and 
$h$ induced by the coupling $\beta_4$, see 
Fig.~\ref{fig2} for the branch (i).
For $n \geq 3$ there is the branch satisfying 
$A_1=\epsilon \sqrt{A_0^2/(fh)}$, but this merely 
corresponds to the RN solutions. 
For the power-law quintic coupling (\ref{G5coupling}), 
it was shown in Sec.~\ref{G5powersec} that the two 
asymptotic solutions around the horizon and 
at spatial infinity are discontinuous due to the divergence 
of $A_1$ at $h=1/(2n+1)$.

In Sec.~\ref{G6powersec} we studied BH solutions for the 
power-law sixth-order coupling (\ref{G6coupling}) and 
found that the consistent branch for $n \geq 0$ corresponds to
$A_1=0$. 
For $n=0$, there exists the $U(1)$ gauge symmetry, 
in which case the integration constant $P$ in Eq.~(\ref{A0P}) does not have physical meaning. 
For $n \geq 1$, 
the Proca hair $P$ is of the secondary type by reflecting the fact that the near-horizon expansions 
(\ref{fn=1})-(\ref{A0n=1}) contain only 
two parameters $\mu$ and $r_h$.
In Fig.~\ref{fig3} the solutions in two asymptotic regimes $r \simeq r_h$ and $r \gg r_h$ smoothly join each other with the largest difference between $f$ and $h$ around the horizon.
Compared to the cases of cubic and quartic couplings plotted 
in Figs.~\ref{fig1} and \ref{fig2}, $|f-h|$ decreases faster 
for increasing $r$. 

In Sec.~\ref{g45powersec} we also discussed the role of 
intrinsic vector modes given by the two power-law 
interactions (\ref{ga45coupling}). 
For the coupling $g_4(X)=\gamma_4 (X/M_{\rm pl}^2)^n$, the branch leading to the difference 
between the two metric components
corresponds to $A_1=0$. 
The coupling $\gamma_4$ induces corrections 
to the RN solutions with the secondary Proca hair $P$ 
arising for $r \to \infty$.
The coupling $g_5(X)=(\gamma_5/M_{\rm pl}^2)(X/M_{\rm pl}^2)^n$ gives rise to the non-vanishing $A_1$ branch (\ref{A1g5}). This is rather a specific case in which $A_1$ approaches finite constants for both the limits 
$r \to r_h$ and $r \to \infty$, with the secondary
hair $P$ at spatial infinity.
The numerical simulation of Fig.~\ref{fig4} for $n=1$ 
shows that, unlike the power-law quintic interaction (\ref{G5coupling}), the solutions are regular throughout the horizon exterior.

In summary, for the power-law models with cubic and quartic 
couplings $G_3(X)$ and $G_4(X)$, we showed the existence 
of regular BH solutions with a primary hair related to the 
longitudinal vector propagation. 
The power-law couplings $G_6(X), g_4(X), g_5(X)$ 
associated with intrinsic vector modes generally give rise 
to regular BH solutions with a secondary hair. 
In both cases the deviation from GR is most significant 
in the strong-gravity regime, with the recovery of 
GR at spatial infinity. The deviation from GR can be potentially probed in future measurements of gravitational 
waves in the nonlinear regime of gravity.

There are several issues we did not address in this paper.
While we focused on the static and spherically symmetric configurations as a first step, 
we can extend the analysis 
to asymptotically non-flat BH solutions and rotating BH solutions, see Refs. \cite{Minami,Cisterna,Babichev17} 
for such solutions in the specific models. 
In particular the existence of hairy Kerr BH 
solutions and stars with gravitational solitons
was recently found for a complex 
Proca field \cite{Herdeiro2,Brito}, 
so it is of interest to study what happens in the presence of vector-field derivative couplings. In addition, the stability analysis of BH solutions against odd- and even-parity perturbations along the line of Refs.~\cite{Suyama} may constrain the strength of derivative couplings studied  in this paper. 
It will also be interesting to investigate solutions of NSs and other (exotic) compact objects
in generalized Proca theories,
where the deviations from GR may be more evident than the BH case,
see Ref. \cite{Chagoya2} for NS solutions in the model with nonminimal coupling to the Einstein tensor.
These interesting issues will be left for future works.

\section*{Appendix: Coefficients in the gravitational 
equations}

In Eqs.~(\ref{be1})-(\ref{be3}) the coefficients 
$c_{1,2,\cdots,19}$ are given by 
\bea
c_{1} &=& -A_{1} X G_{3,X},
\\
c_{2} &=& -2 G_{4} + 4 (X_{0} + 2 X_{1}) G_{4,X} + 8 X_{1} X G_{4,XX},
\\
c_{3} &=& -A_{1} (3 h X_{0} + 5 h X_{1} - X) G_{5,X} - 2 h A_{1} X_{1} X G_{5,XX},
\\
c_{4} &=& G_{2} - 2 X_{0} G_{2,X} - \frac{h}{f} (A_{0} A_{1} A_{0}' + 2 f X A_{1}') G_{3,X} 
-\frac{h A'^{2}_{0}(1+2 G_{2,F})}{2 f}\,,
\\
c_{5} &=& -4 h A_{1} X_{0} G_{3,X} - 4 h^{2} A_{1} A'_{1} G_{4,X} 
+ \frac{8 h}{f} \left( A_{0} X_{1} A'_{0} - f h A_{1} X A'_{1} \right) G_{4,XX}
+\frac{2 h^{2}}{f} A_{1} A'^{2}_{0} (g_{5} + 2 X_{0} g_{5,X}),
\\
c_{6} &=& 2 (1 - h) G_{4} + 4 (h X - X_{0}) G_{4,X} +8 h X_{0} X_{1} G_{4,XX} 
- \frac{h}{f} \left[ (h - 1) A_{0} A_{1} A'_{0} + 2 f (3 h X_{1} + h X_{0} - X) 
A'_{1} \right] G_{5,X}
\\&&-\frac{2 h^{2} X_{1}}{f} (A_{0} A_{1} A'_{0} + 2 f X A'_{1}) G_{5,XX} 
+ \frac{h A'^{2}_{0}}{f} \left[ (h - 1) G_{6} + 2 (h X - X_{0}) G_{6,X} + 4 h X_{0} X_{1} G_{6,XX} \right],
\\
c_{7} &=& -G_{2} + 2 X_{1} G_{2,X} - \frac{h}{f} A_{0} A_{1} A'_{0} G_{3,X}
+\frac{h A'^{2}_{0}(1+2 G_{2,F})}{2 f}\,,
\\
c_{8} &=& 4 h A_{1} X_{1} G_{3,X} + \frac{4 h}{f} A_{0} A'_{0} (G_{4,X}+2 X_{1} G_{4,XX})
-\frac{2 h^{2}}{f} A_{1} A'^{2}_{0} (3 g_{5} + 2 X_{1} g_{5,X})\,, 
\\
c_{9} &=& 2 (h - 1) G_{4} - 4 (2 h - 1) X_{1} G_{4,X} - 8 h X_{1}^{2} G_{4,XX} 
- \frac{h}{f} A_{0} A_{1} A'_{0}\left[ (3 h -1) G_{5,X} + 2 h X_{1} G_{5,XX} \right] 
\\&&-\frac{h}{f} A'^{2}_{0} \left[ (3 h - 1) G_{6} + 2 (6 h - 1) X_{1} G_{6,X} + 4 h X_{1}^{2} G_{6,XX} \right],
\\
c_{10} &=& -\frac{2h}{f} (G_{4} - 2 X G_{4,X})\,,
\\
c_{11} &=& -\frac{2h^{2}}{f} A_{1} X G_{5,X}\,,
\\
c_{12} &=& \frac{h}{f^2} [G_{4} - 2 (2 X_{0} + X_{1}) G_{4,X} - 4 X_{0} X G_{4,XX}]\,,
\\
c_{13} &=& \frac{h^{2}}{f^2} A_{1} [(3 X_{0} + X_{1}) G_{5,X} + 2 X_{0} X G_{5,XX}]\,,
\\
c_{14} &=& - \frac{h}{f} A_{1} [(3 X_{0} + 5 X_{1}) G_{5,X} + 2 X_{1} X G_{5,XX}]\,,
\\
c_{15} &=& \frac{h}{f^{2}}
[ 2 f A_{1} X_{0} G_{3,X} + 2 (2 A_{0} A'_{0} - f h A_{1} A'_{1}) G_{4,X}
+ 4 \left\{ A_{0} (2 X_{0} + X_{1}) A'_{0} - f h A_{1} X A'_{1} \right\} G_{4,XX} \nonumber \\
& &-h A_{1} A'^{2}_{0} (g_{5} + 2 X_{0} g_{5,X})],
\\
c_{16} &=& -\frac{h}{f^{2}}\left[2 f (G_{4} - 2 X G_{4,X} + 4 X_{0} X_{1} G_{4,XX}) 
+ h \left\{ 3 A_{0} A_{1} A'_{0} + 2 f (X_{0} + 3 X_{1}) A'_{1} \right\} G_{5,X}
\right.
\\ && \left.
+2 h \left\{ A_{0} A_{1} (X_{1} + 2 X_{0}) A'_{0} + 2 f X_{1} X A'_{1} \right\} G_{5,XX}
+ h A'^{2}_{0} (G_{6} + 2 X G_{6,X} + 4 X_{0} X_{1} G_{6,XX}) \right],
\\
c_{17} &=& - 2 G_{4} + 8 X_{1} (G_{4,X} + X_{1} G_{4,XX}) 
\\ && + \frac{h A'_{0}}{f} \left[ A_{0} A_{1} (3 G_{5,X} + 2 X_{1} G_{5,XX}) 
+ A'_{0} \left\{ 3 G_{6} + 4 X_{1} (3 G_{6,X} + X_{1} G_{6,XX}) \right\} \right], 
\\
c_{18} &=& 2 G_{2} - \frac{2 h}{f} \left[ (A_{0} A_{1} A'_{0} + 2 f X_{1} A'_{1}) G_{3,X}
+ 2 (A_{0} A''_{0} + A'^{2}_{0}) G_{4,X}
+ 2 A'_{0} (2 X_{0} A'_{0} - h A_{0} A_{1} A'_{1}) G_{4,XX} \right]
\\ &&+ \frac{2 h^{2} A'_{0}}{f^{2}} \left[ f (2 A_{1} A''_{0} + A'_{0} A'_{1}) g_{5}
+ A'_{0} (A_{0} A_{1} A'_{0} + 2 f X_{1} A'_{1}) g_{5,X} \right] 
+\frac{h}{f} A'^{2}_{0}\,,
\\
c_{19} &=& \frac{2 h}{f} \left[ - 2 (A_{0} A'_{0} + f h A_{1} A'_{1}) G_{4,X} 
+ 4 X_{1} (A_{0} A'_{0} - f h A_{1} A'_{1}) G_{4,XX} 
+ h (A_{1} A'^{2}_{0} + A_{0} A'_{0} A'_{1} + A_{0} A_{1} A''_{0}) G_{5,X} 
\right.
\\ && 
+ 2 h A'_{0} (A_{0} X_{1} A'_{1} + A_{1} X_{0} A'_{0}) G_{5,XX}
+ 2 h A'_{0} A''_{0} G_{6}
+ \frac{ h A'_{0}}{f} \left\{ (A_{0} A'^{2}_{0} + 4 f X_{1} A''_{0} - 3 f h A_{1} A'_{0} A'_{1}) G_{6,X}
\right.
\\ && \left.\left.
+ 2 A'_{0} X_{1} (A_{0} A'_{0} - f h A_{1} A'_{1}) G_{6,XX} \right\} \right].
\eea

\section*{Acknowledgements}

LH thanks financial support from Dr.~Max R\"ossler, 
the Walter Haefner Foundation and the ETH Zurich
Foundation.  
RK is supported by the Grant-in-Aid for Young Scientists B of the JSPS No.\,17K14297. 
ST is supported by the Grant-in-Aid for Scientific Research Fund of the JSPS No.~16K05359 and 
MEXT KAKENHI Grant-in-Aid for 
Scientific Research on Innovative Areas ``Cosmic Acceleration'' (No.\,15H05890).
MM was supported by FCT-Portugal through Grant No. SFRH/BPD/88299/2012. 
We also wish to acknowledge the hospitality of the Yukawa Institute for 
Theoretical Physics where part of this work was conducted 
during the workshop on gravity and cosmology for young researchers
(YITP-X-16-10).


\end{document}